# Metaverse for Industry 5.0 in NextG Communications: Potential Applications and Future Challenges

B. Prabadevi, N. Deepa, Nancy Victor, Thippa Reddy Gadekallu, Praveen Kumar Reddy Maddikunta, Gokul Yenduri, Wei Wang, Quoc Viet Pham, Thien Huynh-The, and Madhusanka Liyanage

*Abstract*—With the advent of new technologies and endeavors for automation in almost all day-to-day activities, the recent discussions on the metaverse life have a greater expectation. Furthermore, we are in the era of the fifth industrial revolution, where machines and humans collaborate to maximize productivity with the effective utilization of human intelligence and other resources. Hence, Industry 5.0 in the metaverse may have tremendous technological integration for a more immersive experience and enhanced communication.These technological amalgamations are suitable for the present environment and entirely different from the previous perception of virtual technologies. This work presents a comprehensive review of the applications of the metaverse in Industry 5.0 (so-called industrial metaverse). In particular, we first provide a preliminary to the metaverse and industry 5.0 and discuss key enabling technologies of the industrial metaverse, including virtual and augmented reality, 3D modeling, artificial intelligence, edge computing, digital twin, blockchain, and 6G communication networks. This work then explores diverse metaverse applications in Industry 5.0 vertical domains like Society 5.0, agriculture, supply chain management, healthcare, education, and transportation. A number of research projects are presented to showcase the conceptualization and implementation of the industrial metaverse. Furthermore, various challenges in realizing the industrial metaverse, feasible solutions, and future directions for further research have been presented.

B. Prabadevi, Nancy Victor, Praveen Kumar Reddy Maddikunta and Gokul Yenduri are with the School of Information Technology and Engineering, Vellore Institute of Technology, Vellore, Tamil Nadu- 632014, India (Email:{prabadevi.b, nancyvictor, thippareddy.g, praveenkumarreddy, gokul.yenduri}@vit.ac.in).

N. Deepa is with the Department of Computer Science, School of Engineering and Technology, Pondicherry University, Karaikal Campus, Karaikal - 609605, Puducherry Union Territory, India (email: deesraj@pondiuni.ac.in).

Thippa Reddy Gadekallu is with the Zhongda Group, Haiyan County, Jiaxing City, Zhejiang Province, China, 314312, as well as with the Department of Electrical and Computer Engineering, Lebanese American University, Byblos, Lebanon, as well as with the School of Information Technology and Engineering, Vellore Institute of Technology, India, as well as with the College of Information Science and Engineering, Jiaxing University , Jiaxing 314001,China and with the Division of Research and development, Lovely Professional University, Phagwara, India(e-mail: thippareddy@ieee.org)

Wei Wang is with the School of Medical Technology, Beijing Institute of Technology, Beijing, China, and Department of Engineering, Shenzhen MSU-BIT University, Shenzhen, China. Email: ehomewang@ieee.org

Quoc Viet Pham is with the School of Computer Science and Statistics, Trinity College Dublin, The University of Dublin, Dublin, D02PN40, Ireland (e-mail: viet.pham@tcd.ie).

Thien Huynh-The is with Department of Computer and Communication Engineering, Ho Chi Minh City University of Technology and Education, Vietnam (email: thienht@hcmute.edu.vn).

Madhusanka Liyanage is with the School of Computer Science, UCD, Ireland (e-mail:madhusanka@ucd.ie).

## I. INTRODUCTION

The current industrial revolution, Industry 4.0, revolutionized manufacturing and allied sectors by bringing disruptive technologies such as cognitive computing, artificial intelligence (AI), cloud computing, and cyber-physical systems (CPS) to the forefront of manufacturing. Industry 4.0 allowed the machines to be intelligent, communicate with each other, and do much of the production work in factories, giving us the term "smart factories" [1]. "Mass personalization" is enabled by Industry 4.0, where customers can personalize the products they want to purchase online through mass personalization techniques. Mass personalization is realized through the Internet connectivity between the supply chains, robots involved in the manufacturing process, dealership ordering systems, and the customers [2]. While Industry 4.0 focused on the connectivity of CPS, the upcoming Industry 5.0 revolves around the relationship between man and machine. Industry 5.0 promises to integrate the creativity of humans with the precision of robots. The machinery enabled by digital technologies combined with the cognitive intelligence of humans is expected to enhance the communication and also increase the speed of production/manufacturing Table II represents the list of acronyms used in this work.

As many Industry 5.0 applications are mission-critical and require real-time decisions, a platform that helps human experts in having access to the immersive experience of the situation before making a decision is mandated. It has immense potential in minimizing the losses in property and lives and provides customers better products. The metaverse is a perfect fit to bridge this gap in potential Industry 5.0 applications. The metaverse has been in the limelight recently with the claims from Microsoft and Facebook. The metaverse combines several technologies, such as video, augmented reality (AR), and virtual reality (VR), in which users can work, play with friends, and stay connected with their friends virtually through conferences, virtual trips, and concerts. The metaverse's expansion will allow humans to co-exist in a hyper-real alternate world [3]. Facebook is investing heavily in this technology as it envisions and foresees a virtual world in which digital avatars are connected through entertainment, travel, or work using VR headsets. Facebook believes that the metaverse could replace the Internet. The metaverse is expected to make the Internet medium more embodied and immersive, where people not just look at the Internet but also would experience it [4]. With its exciting features, the metaverse has immense potential to take the upcoming and futuristic Industry 5.0 to a new level. In the Metaverse, virtual replicas of the products in manufacturing can be created in the digital world, where experts can see the progress in manufacturing virtually, patients can be treated in the digital world by doctors virtually, and people can meet their peers and play exciting video games in the digital world, organizations can track the products throughout their supply chain life-cycle in the digital world, the governments can plan the infrastructure in the smart city [5], [6] by visualizing the smart city projects in the digital world and also they can be better equipped to deal and respond to the natural disasters, people can create digital assets, experience the products and also can do the shopping by experiencing the immersive technology and the list goes on. In this study, we aim to present a comprehensive review of the applications of the metaverse in the realization of the true potential of Industry 5.0.

Due to the immense potential of Industry 5.0 and the metaverse in revolutionizing the industry and people's lifestyles, several researchers recently published surveys on both these technologies. In [7], the authors considered some of significant enabling technologies for Industry 5.0, such as the Internet of Things (IoT), Artificial Intelligence (AI), Augmented Reality (AR), Virtual Reality (VR), Big Data Analytics, Edge Computing, and 5G. The authors also presented industry 5.0 applications such as healthcare, Supply Chain Management (SCM), Smart Education, Cloud Manufacturing. The study in [8] presented a survey on applications of Industry 5.0 for COVID-19. The authors reviewed how humans and robots work together to perform jobs like surgery, treatment, and monitoring patients. This helps doctors and nurses to provide personalized care for patients. The collaboration among humans and machines in healthcare develop the quality of healthcare and patient outcomes. Another motivating study in [9] discussed key components of Industry 5.0 from a manufacturing perspective. The authors examined AI, robotics, intelligent machines, IoT, and big data analytics as significant elements of Industry 5.0 in the manufacturing industry. These technologies can increase throughput, effectiveness, quality, and cost reduction. The authors in [10] presented a detailed survey on how humans and robots can collaborate. The authors discuss how human-robot collaboration has the potential to benefit businesses and employees. The authors highlight some organizational challenges such as safety, cost, and change management, as well as human employee issues such as job security, trust, and social interaction. It is necessary to address all these issues so that organizations can achieve the benefits of human-robot collaboration. Another interesting study in [11] discussed how Industry 5.0 could impact education in engineering courses. According to the authors, the current engineering curriculum is insufficient to prepare engineers for the challenges of Industry 5.0. They propose four ways to improve engineering education: prioritize lifetime learning, prioritize transdisciplinary education, emphasize soft skills, and use active learning approaches. Following these ideas in engineering education prepares students for the challenges of Industry 5.0. The study in [12] presented a detailed discussion on how Industry 5.0 can revolutionize the manufacturing sector in the pharmaceutical industry. The authors identify several challenges that pharmaceutical companies must overcome in order to fully implement Industry 5.0. The authors provide solutions for pharmaceutical companies to overcome the challenges of meeting Industry 5.0 standards. In another interesting study, the authors in [13] presented a comprehensive survey on human-centric manufacturing in Industry 5.0. The authors highlight the importance of humans in the manufacturing process. Machines can help to automate processes, but they cannot replace the human touch. Humans are still required to make decisions, solve problems, and interact with customers. Human-centric manufacturing places humans at the center of manufacturing decision-making. Multiple technologies are used in this approach to improve the work environment and make it more efficient and productive. On the other hand, the study in [14] presented a comprehensive survey on the factors that support a viable and functional metaverse. The authors discussed how advances in hardware performance, such as faster CPUs and better graphics cards, allow users to have more immersive experiences. Connectivity and internet infrastructure are important for maintaining stable user communication. Advanced computer approaches and algorithms are required to create realistic simulations. User participation, user adoption, and content creation all contribute to the growth of the metaverse. Finally, the authors suggest that addressing regulatory and ethical concerns is important for maintaining a safe and trustworthy virtual environment. The study in [4] presented a detailed survey on the taxonomy, key components, potential applications, and open challenges of the metaverse. Even though several researchers presented review papers on Industry 5.0 and the metaverse separately, very few reviews exist on the fusion of these two exciting technologies. Table I explains the summary of related reviews on Industry 5.0 and the metaverse.

The above observations have motivated us to conduct this comprehensive review on the applications of the metaverse for Industry 5.0. The main contributions of this study are:

- We provide the definitions of the metaverse and enabling technologies of the metaverse and Industry 5.0.
- The potential applications of the metaverse in several Industry 5.0 applications are presented.
- Some of the key research and industry projects related to the applications of the metaverse in several Industry 5.0 verticals are discussed.
- Several challenges that may arise in the fusion of the metaverse with Industry 5.0 applications are discussed. We also provide future research opportunities that drive the researchers and industry towards future research in this interesting fusion.

The rest of the paper is organized as follows. Some of the important definitions and key technologies of the metaverse and Industry 5.0 along with the motivation for the fusion of the metaverse with Industry 5.0 are discussed in Section 2. The potential applications of the metaverse in different verticals of Industry 5.0 are presented in Section 3. Section 4 discusses some of the important research and industry projects that are focused on the fusion of the metaverse and Industry 5.0. The key challenges, open issues and future research directions towards the fusion of the metaverse and Industry 5.0 are discussed in Section 5. Finally, we conclude our study in Section 6.

TABLE I
SUMMARY OF RELATED REVIEW PAPERS.

| Ref. | Contributions | Limitations |
|---|---|---|
| [7] | Presented a comprehensive survey on the key enabling technologies and potential applications of Industry 5.0. | The applications of the metaverse in Industry 5.0 have not been presented. |
| [10] | Presented a detailed survey on how humans and robots can collaborate together. | The discussion of the metaverse in Industry 5.0 has not been provided. |
| [4] | Presented a detailed survey on the taxonomy, key components, potential applications, and open challenges of the metaverse. | This study focused on potential applications, and open challenges of the metaverse and didn't consider the integration of the metaverse in Industry 5.0. |
| [8] | Presented a detailed survey on potential applications of Industry 5.0 for COVID-19 | The authors only focused on a subset of healthcare and integration of the metaverse for COVID-19 remains untouched which is a serious limitation of this work. |
| [9] | Discussed key and critical components of Industry 5.0 from a manufacturing perspective. | The authors present key and critical components of Industry 5.0 from a manufacturing perspective and did not focus on the integration of the metaverse in Industry 5.0. |
| [11] | Discussed how Industry 5.0 can impact the education in engineering courses. | Their study limits itself to the integration of metaverse in education. |
| [12] | Presented a detailed discussion on how Industry 5.0 can revolutionize the manufacturing sector in the pharmaceutical industry. | This study focused on potential applications, and open challenges of the metaverse in the pharmaceutical industry and The study hasn't discussed the role of the metaverse in the pharmaceutical industry. |
| [13] | Presented a comprehensive survey on human-centric manufacturing in Industry 5.0. | This study focused on human-centric manufacturing in Industry 5.0 and the study did not focus on the integration of the metaverse in human-centric manufacturing in Industry 5.0 [15]. |
| [14] | Presented a comprehensive survey on the factors that support a viable and functional metaverse. | The authors have not discussed the applications of the metaverse for Industry 5.0. |
| This paper | This work presents a comprehensive review of the applications of the metaverse in Industry 5.0. | |

TABLE II
LIST OF KEY ACRONYMS.

| Acronyms | Description |
|---|---|
| AI | Artificial Intelligence |
| AR | Augmented Reality |
| CPS | Cyber-physical Systems |
| CMfg | Cloud Manufacturing |
| DT | Digital Twin |
| FL | Federated Learning |
| HCI | Human Computer Interface |
| ICT | Information and Communication Technologies |
| IoT | Internet of things |
| MR | Mixed Reality |
| NFT | Non-Fungible Token |
| QC | Quantum Computing |
| SCM | Supply chain management |
| VR | Virtual Reality |
| XR | Extended Reality |
| 3D | Three-Dimensional |
| 3GPP | 3rd Generation Partnership Project organization |
| 6G | Sixth Generation |

## II. PRELIMINARIES OF THE INDUSTRIAL METAVERSE

This section presents a brief overview of the metaverse and enabling technologies of the industrial metaverse.

### A. The Metaverse

Neil Stevenson coined the term "metaverse" in his 1992 novel Snow Crash [4]. However, it attracted huge attention in October 2021, when Facebook officially changed its name to Meta [16], [17]. A fully functioning persistent metaverse does not yet exist. However, there are metaverse-like platforms such as Roblox, Decentraland, Axie infinity, Illuvium, Sandbox Fortnight, and SecondLife, etc [15]. The metaverse is a shared virtual environment for multiple users that combines physical reality with the digital virtual world [18]. It is based on the fusion of cutting-edge technologies that enable multidimensional interactions. Hence, the metaverse is a platform of interconnected environments. It allows users to communicate with each other in real time and interact with digital artifacts. The metaverse will significantly impact society, agriculture, education, healthcare, and other sectors. AI, IoT, blockchain, digital twins, and AR will all be able to reach their full potential in enabling the metaverse.

### B. Definitions

**Definition 1:** "The metaverse is a digital reality that combines aspects of social media, online gaming, AR, VR, and cryptocurrencies to allow users to interact virtually [19]."

**Definition 2:** "The metaverse is a network of 3D virtual worlds focused on social connection. In futurism and science fiction, the term is often described as a hypothetical iteration of the Internet as a single, universal virtual world facilitated by virtual and augmented reality headsets [20]."

**Definition 3:** "The word "metaverse" is a portmanteau of the prefix "meta" (meaning beyond) and "universe"; the term is typically used to describe the concept of a future generation of the internet, made up of persistent, shared, 3D virtual spaces linked into a perceived virtual universe [21]."

**Definition 4:** "The metaverse is a persistent and immersive simulated world experienced in the first person who shares a strong sense of mutual presence. It can be fully virtual (i.e., a virtual metaverse), or it can exist as layers of virtual content overlaid on the real world (i.e., an augmented metaverse) [22]."

**Definition 5:** "The metaverse: a persistent, live digital universe that affords individuals a sense of agency, social presence, and shared spatial awareness, along with the ability to participate in an extensive virtual economy with profound societal impact [23]."

**Definition 6:** "An image of virtual everything: You attend work meetings as an avatar using the Quest VR headset and use a device on your wrist to secretly text friends [24]."

**Definition 7:** "The metaverse is the sum of virtual and augmented realities concentrated on a super long "Street" through which people walk as avatars and can access using goggles and plugging into terminals [24]."

**Definition 8:** "The convergence of virtually-enhanced physical reality and a physically persistent virtual space [25]."

**Definition 9:** "The metaverse is an expansive network of persistent, real-time rendered 3D worlds and simulations [26]."

### C. Key Enabling Technologies of the Industrial Metaverse

The metaverse is a virtual world where users can socialise, do business, and play video games. As more individuals participate, the metaverse's popularity will grow exponentially. In addition to Meta, Microsoft, and Apple, other technology companies have begun investing in the metaverse. New users are attracted by cryptocurrencies, NFTs, and play-to-earn games. The metaverse is significant as it represents the future of social media. The metaverse is predicted to become a worldwide phenomenon because of its novel 3D world, socialising, and gaming capabilities.

The realization of the industrial metaverse is feasible by the key enabling technologies depicted in Fig. 1. In this section, the key enabling technologies that can help the realization of the potential of the metaverse in Industry 5.0 are presented.

*1) Virtual Reality:* VR is a 3D computer-rendered virtual environment. VR makes individuals feel immersed with surrounding scenes and objects as if one of the characters of the scene appears to be real [27]. The VR headset or helmet will be used to perceive the simulated environment allowing users to interact with the virtual objects through user interaction techniques such as head tracking or tangible controllers. Since it is separated from physical reality the users with the gadgets need to be more focused in the virtual environment. The recent advancements in VR led to the basis for the innovation of the metaverse. VR and the metaverse are not the same. However, VR is an element of the metaverse, which provides a virtual space for implanting multiple users by enabling interactions simultaneously [28]. In Industry 5.0, the metaverse will deliver humans the best virtual experiences. Industry 5.0 enables the collaboration between machines and humans in developing customized products. The metaverse with the assistance of VR will remove the barriers between reality and the virtual world [29]. The metaverse will allow the user to personalize the products using VR. Industry 5.0 with the blend of advanced technologies or tools like VR can produce efficient products with all specifications of the customer and also contribute to the business by incorporating smart automation, creativity, and problem-solving skills of the human/robot partners.

*2) 3D Modeling:* 3D modeling is a process of creating a 3D representation of any object or surface, which helps in creating 3D virtual environments. Every physical element will have a digitized aspect to it. Converting it into 3D elements will improve the essence of the representing real world in a virtual environment [30]. The metaverse is a shared virtual 3D environment which is interactive, immersive, and collaborative. 3D modeling is a key component of the metaverse which helps in creating 3D avatars and 3D spaces enabling users to interact with each other or to perform operations in a virtual world. The metaverse is totally dependent on 3D captures and visualization. Computing devices need to understand the visual information of the user activities and their surroundings which helps in building a realistic 3D Virtual environment in the metaverse. Based on the activities of the user, automatic reconstruction of the 3D virtual environment will also take place without interrupting the operations, using efficient body and pose tracking algorithms in the metaverse. In Industry 5.0, it is essential to implement new methods to satisfy and understand customer needs and then into products and services. The recent advancement in 3D modeling will help in visualizing the transformation of the raw materials into finished goods. The metaverse with modern technologies will lead to mass customization of products by creating a relationship between humans and machines in Industry 5.0 [31].

*3) Artificial Intelligence:* Recently, several organizations started using AI for their business operations. AI is the ability of a computer or computer-controlled robots to perform the tasks like humans [32]. Users can use AI for decision-making or automation process. AI will assist the metaverse by including content analysis, supervised speech processing, and computer vision [33], [34]. AI also helps in transforming the role of entities from the physical world to the virtual environment automatically. The AI can help the metaverse in different ways.

1) AI can analyze the user images to create more realistic avatars with a variety of facial expressions, emotions, hairstyles, features brought on by aging, etc.
2) For creating AI-enabled nonplaying characters i.e., digital humans, to respond to the actions of avatars in a VR world.
3) Natural language understanding (NLU) and natural language generating (NLG) capabilities of AI enable seamless and multilingual communication in the metaverse.
4) Using AI techniques, the metaverse can be extended without the intervention of humans. AI can also assist in human-computer interactions.

Automation and personalization of various services will help Industry 5.0 to enhance the user's experience. Mass customization of the products increases productivity by improving the efficiency of the collaborators between humans and machines using AI-based methods, such as machine learning, deep learning, convolutional neural networks, recurrent neural networks, reinforcement learning, generative adversarial networks, stacked autoencoders, graph neural networks, and meta-heuristic algorithms. AI-based techniques, provide a basic foundation for Industry 5.0 that can help in understanding and enhancing the requirements by enabling the concepts of Industry 5.0.

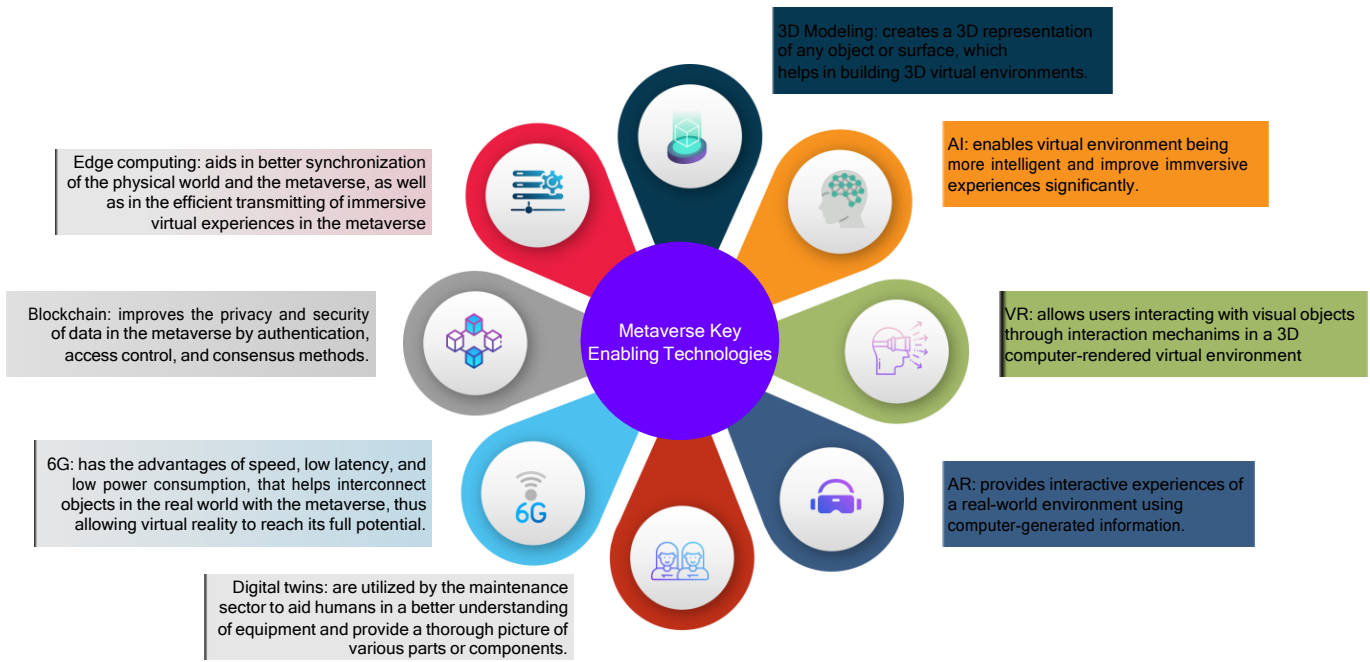

Fig. 1. The key enabling technologies of the industrial metaverse.

*4) Augmented Reality:* AR is a technology that provides an interactive experience of a real-world environment using computer-generated information. Superimposing virtual objects into a real world based on the user's perspective will enhance the effectiveness of the users' interaction [35]. AR provides more realistic solutions by utilizing minimal hardware and also provides a solid foundation for the development of the metaverse. AR object detection techniques will help to build an efficient immersive 3D environment in the metaverse. Along with the existing technologies, the usage of modern technologies, such as blockchain in the metaverse will change the user's perspective of interacting with the Internet by launching the virtual ecosystem for various applications like virtual office space. The metaverse also opens a new opportunity for tourism, education, entertainment, retail, engineering, and many more industries [36]. In Industry 5.0, the main focus is on mass customization and ultra-personalization, which requires the blend of the latest technologies like AR to establish collaboration between cognitive systems, robots, and humans [11]. AR connects the virtual world and the physical world by overlaying the data with the physical world. AR in Industry 5.0 will guide the technicians or workers in machine maintenance i.e., performing service in real-time, and also the digital management of the workspace can be achieved by maintaining the harmony between the business and manufacturing process.

*5) Blockchain:* In 2008, Nakamoto Satoshi published a white paper that served as the foundation for the blockchain concept. A blockchain, also known as a distributed ledger, is made up of a series of connected blocks, each of which is linked to the preceding one using the hash value of its header [37]. A block also contains a timestamp, nonce, and transaction data in addition to the cryptographic hash. All nodes must agree to the same consensus process in order to generate and validate new blocks of data. The data exchange in the metaverse is greatly aided by using the advanced encoding information system of the blockchain [38]. The privacy of sensitive information is protected by the blockchain's authentication, access control, and consensus methods [39]. Individuals and organizations can verify all transactions due to the detailed audit trail provided by the blockchain. The quality of data in the metaverse will improve as a result of integrating blockchain with the metaverse. The interaction between the users in several applications of Industry 5.0 will lead to the exchange of sensitive information [40]. Multiple parties may also be involved in this transit. As a consequence, users need to have greater control over their data in Industry 5.0, so that their sensitive information does not leak into the wrong hands. In this case, Industry 5.0 will effectively deliver the services with the help of the metaverse. The metaverse will safeguard the integrity and transparency of information through the transactions in a blockchain. This also ensures users' data privacy in Industry 5.0. Data stored in different applications of Industry 5.0 is very important and attackers will try to exploit the storage. If the data stored is tampered with, it will directly affect the outcomes of the applications, which may lead to unnecessary complications. On the other hand, the metaverse will allow users to interact with organizations virtually and allow them to customize products [41]. The blockchain will protect this sensitive data with its consensus mechanism, which will only allow the modification of stored data based on the agreement of all the participating nodes in the blockchain [42]. This makes the data storage in Industry 5.0 resistant to attacks. Man and machine collaboration requires secure data transfer across various applications of Industry 5.0. For example, in smart manufacturing, the user may require to customize a product according to his choice and tries to send the data to the manufacturer. There is a chance that the data related to the customized design can be modified by the intruder in this process of data exchange. In this case, the metaverse will provide a virtual space for the user to create a design of their choice and will also share the same with the organization with the help of the advanced encoding system provided by the blockchain, which makes Industry 5.0 data exchange secure. The integration of blockchain and the metaverse will enable Industry 5.0 and its users with data privacy for their shared information, trust in the data stored, and also provide reliability in data exchange.

*6) 6G:* 6G is the sixth-generation standard for wireless communications technology that is currently in development. It is the successor to 5G and will be significantly faster. 6G networks will be even more heterogeneous than their predecessors and support VR/AR, real-time communications, ubiquitous intelligence, and the IoT [43]. Networks and communications are integral to everything from large-scale computation to enabling shared experiences among people. The metaverse will be greatly benefited by 6G networks. 6G has the advantages of speed, low latency, and low power consumption [44]. This will help interconnect objects in the real world with the metaverse. 6G will allow VR to reach its full potential [45], [46]. This will expand cooperation between man and machine by connecting the physical world with the virtual world, establishing the foundation for the metaverse. Numerous applications within Industry 5.0 will undergo a transformation as a result of 6G and the metaverse integration. Through the use of cutting-edge technologies like XR applications, the metaverse enables customers to do online shopping, professionals to attend meetings in their desired avatar, and will also allow students to learn about historical places virtually, with uninterruptible services provided by 6G [47]. The metaverse will enable high-quality services and experiences for users of Industry 5.0, where users can parallelly customize a product while interacting with the people of that particular organization. 6G will enable the metaverse with latency-free service where industry personnel can use holographic communication to discuss effective strategies with their peers in developing products. The integration of 6G and the metaverse will enable Industry 5.0 and its users to use high-quality services and experiences, parallel working capabilities, and latency-free communications.

*7) Edge Computing:* Client-server communication is accelerated with the help of edge computing, resulting in reduced latency and less consumption of bandwidth [48]–[50]. Due to the distributed structure of edge networks, it is difficult to attack. If a breach occurs in a minor section of the network, the rest of the network will not be exposed. The edge computing architecture distributes processing tasks across the network, making it more resilient than other centralized systems [51]. Edge computing will aid in better synchronization of the physical world and the metaverse, as well as in the efficient transmitting of immersive virtual experiences in the metaverse

[52]. Sensor data is critical in making real-time decisions, and edge computing will contribute to it by supplying uninterrupted and faster data services, allowing machines to function more efficiently in collaboration with humans [53]. In the metaverse, edge computing will also ensure reliable resource allocation, which will assist humans in making better decisions based on the outcomes of AI models. Industry 5.0 will greatly benefit from the integration of the metaverse and edge computing. The AI and XR technologies, which are supported by the metaverse, will help Industry 5.0 users make effective decisions, customize designs, and perform tasks from a remote location in real-time. The integration of edge computing and the metaverse will enable Industry 5.0 and its users to make effective decisions, customize designs, and perform tasks from a remote location.

*8) Digital Twins:* A digital twin is a digital representation of an entity or system in the physical world. A digital twin consists virtual object or model that represents a real-world object, process, organization, person, or other abstraction [54]–[56]. In the metaverse, data from various digital twins is pooled for a holistic view of real-world entities and their associated operations. In Industry 5.0, digital twins are utilized by the maintenance sector to aid humans in better understanding of equipment and provide a thorough picture of various parts or components. A holistic view of machines helps organizations save time on machine maintenance by reducing the number of repairs and making it easier for humans to fix machines when they break down. Humans will have the necessary information about the machines in the metaverse with the help of digital twins, which will allow them to make better decisions in Industry 5.0.

### D. Motivation behind the Integration of the Metaverse with Industry 5.0

Industry 4.0 has led the manufacturing industries to focus on utilizing AI/ML algorithms to make effective predictive maintenance, production, and quality decisions. Furthermore, the advent of Industrial Information Integration Engineering integrates various methods from relevant disciplines like ICTs into industrial sectors to enable better industrial processes [57]. Chen and Yong have conducted a systematic review from 2006 to 2015 on how Industrial Information Integration Engineering effectively adopts different perspectives of various domains and usage of business process management, service-oriented architectures, and applications in the enterprise system and research opportunities [58]. They have also presented a systematic literature of industrial information integration for the period of 2016-2019 [59] which emphasizes mainly the integration of IoT, smart grids, CPS, and smart manufacturing with Industrial Information Integration Engineering. Various industrial informatics such as enterprise application integration, SOA, and grid computing, their usage in industrial information integration, and challenges were explored for better adoption [60]. The industrial metaverse can automate the entire environment, inclusive of the full supply chain, factory floors, production, retail and physical assets by replicating it digitally for an efficient remote view [61]. The problems at the factory floor or retail can be better viewed without physical presence. The advantage of the metaverse in Industry 4.0 is cost optimization as it provisions virtual prototyping, letting companies to create virtual representations of items, manufacturing facilities, and production, enabling efficient cost optimization without the necessity for physical resources. In Industry 4.0, the metaverse supports the virtual simulation and analysis of designs, hence lowering risks. In Industry 4.0, the metaverse provides a virtual interface for remote monitoring, allowing technicians to diagnose equipment issues in real-time.

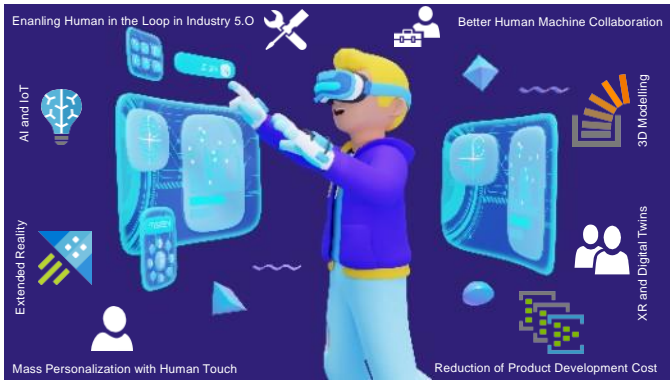

Fig. 2. Motivation for Integration of the Metaverse with Industry 5.0.

Industry 5.0 is the next generation of the industrial revolution and promises to integrate human creativity with efficient, intelligent, and precise technology of machines in order to create resource-efficient and user-friendly solutions [35]. Due to this new standard greater efficiency and adaptability can be achieved. Humans and robots can collaborate more efficiently and creatively in Industry 5.0. IoT, AI, digital twins, big data, and robotics will play a significant role in Industry 5.0, which assists humans in developing society, agriculture, education, healthcare, and other sectors [13]. In Industry 5.0, humans and machines work together to develop high-quality, high-speed products. The metaverse integration with Industry 5.0 will attract more customers through virtual factory tours and interactive sessions, which are enabled with the help of XR devices. The metaverse also allows virtual product presentations and aids in mass personalization. Payments can be securely processed between organizations and users through the use of blockchain technologies. AI-powered digital assistants are used in the metaverse which will improve customer service and the real-world customer experience.

The benefits of integrating the metaverse in Industry 5.0 over Industry 4.0, which motivated us to perform this study, are shown in Fig. 2.

*1) Better Human-Machine Collaboration:* Machines process information far faster than humans and can conduct essential computations with high accuracy and speed. Collaboration between humans and machines will result in outstanding outcomes in a variety of Industry 5.0 applications. Even the smallest inaccuracy can have catastrophic consequences when working with massive machinery. In this situation, the metaverse and its enabling technologies will provide humans with a virtual arena in which they can simulate tasks using XR applications and analyze the success rate and potential hazards using AI prior to executing the actual work. As a result, the margin for error will be kept to a minimum [62].

*2) Enabling Human in the Loop in Industry 5.0:* In Industry 4.0, automated equipment and AI models are employed to develop products. AI models and automated robots may not have the same level of understanding of experience and emotions as humans. In Industry 5.0, by integrating a variety of sensors, the metaverse will enable people to experience and interact with products in ways that were previously impossible with existing technologies [63]. As a result, humans and machines will work together in Industry 5.0, with the help of the metaverse, to improve the quality of the product. For Example, medical students can benefit from the metaverse in their Education 5.0 by visiting a virtual anatomy lab and performing surgery on virtual patients in collaboration with trained professionals.

*3) Reduction of Product Development Cost:* In order to develop a product effectively and with the highest quality, the designers need accurate information about the product from the client. The client may not be able to provide accurate information about the product's description and specifications. In this scenario, the organizations will have to construct a prototype with the acceptance of the customer. Then the mass production of products will commence if the client is satisfied. This conventional procedure requires a lot of time, manpower, and cost even before the start of large-scale production. In this scenario, the metaverse, with its supporting technologies like XR and digital twins, can help in reducing the stress on real-world systems and assist the humans in making better decisions [64]. This strategy achieves maximum creativity and innovation in designing products in Industry 5.0.

*4) Mass Personalization with Human Touch:* Traditionally, online customers typically rely on the product's visuals and description to make their purchase decisions. The customer will only be able to see the product after it has been delivered and will not be able to make any customization. The customer's dissatisfaction with the quality, design, and delay in delivery may prompt the customer to reject the product. This will result in a loss for the organization if it happens with several customers. Customers in the metaverse will be able to see in 3D how products are created, delivered, and sold in Industry 5.0's supply chain. Due to the improved transparency, customers will be able to see exactly how long it will take for products to arrive and how much it will cost to deliver them. Additionally, clients will be able to personalize their products through direct interaction with the machine [65]. If the customer delivers more detailed design of the product, the error margin will be greatly reduced. This will substantially reduce consumer turnover and product return costs, which will benefit organizations and the customer.

## III. THE METAVERSE FOR DIFFERENT VERTICALS OF INDUSTRY 5.0

The metaverse is a virtual environment that includes computer-generated 3-D objects, avatars, and communication devices. Users from remote locations interact through this environment for a specific purpose in real time. They have been used in several domains from education to entertainment. These domains started exploring the technology for different needs such as providing actual data in the form of 3D maps which will let the user experience knowledge, that cannot be replicated by any other models. The governments have started using the metaverse to discover a virtual model for weather conditions and natural calamities integrated with real-world data. Also, this technology has been applied in various applications of Industry 5.0, which are discussed below. This section explains how the metaverse is used in multiple Industry 5.0 applications.

### A. The Metaverse in Society 5.0

Several prototypes are being developed nowadays to integrate physical hardware devices with virtual worlds. The metaverse can play a significant role in the education sector. The facilities provided by the metaverse for education include research collaboration and meetings to provide various services to the students in a virtual campus environment. This technology can provide a platform for students and faculty to collaborate in a virtual environment for various requirements [66], [67]. Some of the benefits of the metaverse in Society 5.0 include increased communication and collaboration among humans from all over the world via virtual meetings and collaborative workspaces. Through virtual parties, gatherings, and other celebrations, the metaverse has the potential to socialize and connect with others [68]. Society 5.0 has several applications in the metaverse such as robot-assisted exam conduction, an assortment of energy for various purposes, and robotizing the agriculture sector which leads to an increase in production and advancement in sustainable industrial development [69]. Recently Japanese government has proposed a new research policy namely Society 5.0. This is defined as a humanistic society that helps to balance economic development with the intention to solve various social issues by developing a system that can integrate the physical environment with cyberspace. Human civilization is categorized into five phases by the Japanese government. The community of hunting, in which humans lived by hunting animals and gathering plants for their livelihood is called Society 1.0. The second phase is said to be an agrarian society, in which the economy of the society is based on agriculture, namely, society 2.0. The third phase is Society 3.0, in which the agrarian society is transformed into an industrial society. The fourth phase is society 4.0, in which information has become an important resource, called, information society. In Society 4.0, information is not shared and there is no mutual collaboration among various disciplines. This was solved by the Japanese government with the help of IoT [70], an emerging concept, called, Society 5.0. Society 5.0 aims to solve various social problems in the domain such as food, manufacturing, agriculture, healthcare, disaster management, and education by providing digital solutions with the help of advanced technologies such as AI, robotics, big data, and IoT [71].

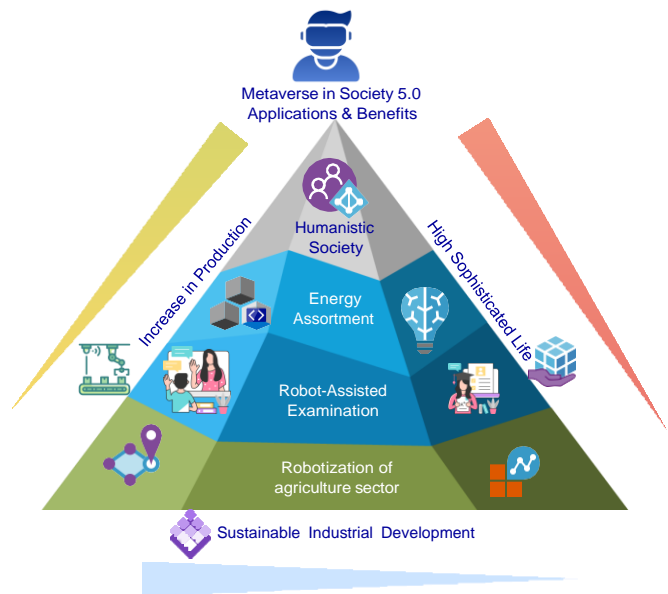

Fig. 3. The metaverse in Society 5.0

One of the remarkable education networks of Japan, namely, KOSEN proposed a sharing system for scientific devices and information using Society 5.0. This education network has more than 50 colleges all over the country. They wanted to share the information and merge their physical infrastructural facilities among their KOSEN colleges. In order to facilitate research collaboration among the physical laboratory areas such as chemistry and material science, they have proposed a remote sharing system for scientific apparatus. Researchers should share the physical experimental equipment in real-time in order to use their skills and share their knowledge in the appropriate domain. In the proposed system, a virtual classroom was created for researchers, in which a general presentation was given to explain the details of the apparatus. Participants share their views through text-based messages and verbally. In order to have a realistic experience, classroom facilities with avatar movement were created to enhance the learning environment. Also, virtual objects representing the scientific equipment were created which resemble the physical objects. The researchers are allowed to use the apparatus remotely and perform the experiments in the metaverse environment [72]. Society 5.0 is viewed as a humanistic society where people can enjoy high sophisticated life. Along with technological innovations, there are a few limitations in the implementation of Society 5.0 in the metaverse environment. The implementation task is complex and expensive due to the utilization of VR and AR equipment. The technology is integrated with various IoT devices, which may lead to the production of a huge volume of confidential data that may lead to security breaches. Enhanced solutions are required to prevent cyber attacks [73]. Robotization in Society 5.0 may affect the employment of human manpower [74]. Fig. 3 depicts the metaverse in society 5.0.

### B. The Metaverse in Agriculture 5.0

Conventional farming practices are being improved by the latest scalable technological advancements such as IoT, AI,

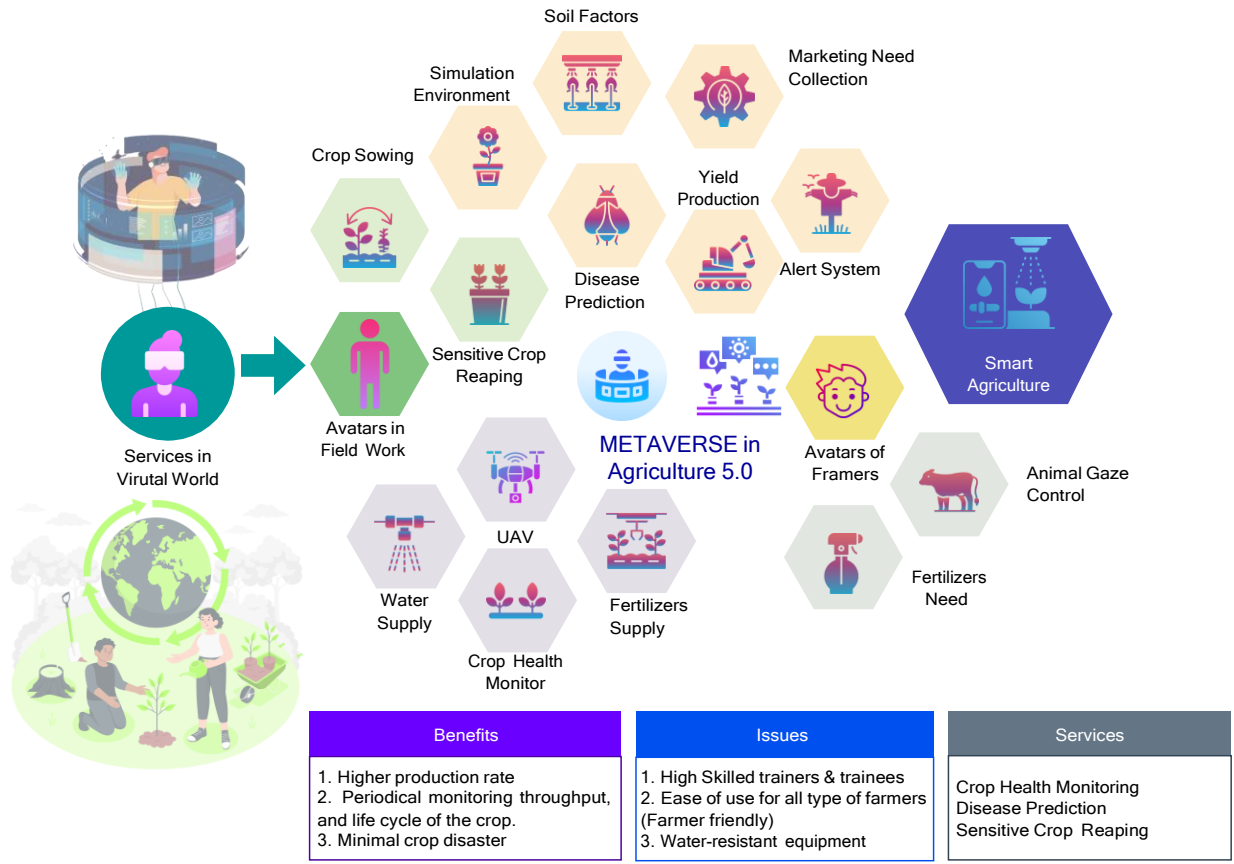

Fig. 4. The metaverse in Agriculture 5.0

and machine learning (ML) which helps to reduce the risks, deliver intelligent decisions, and improve sustainability. Agriculture 5.0 is one of the applications of Industry 5.0, in which smart farms implement precision agriculture and utilize devices that include automated intelligent systems and unmanned vehicles. Agriculture 5.0 also utilizes robots, along with AI techniques. Since the conventional farming system faced a shortage of manpower, Agriculture 5.0 came into existence by integrating agricultural robots with AI techniques. The robots can reap crops in huge amounts that are significantly faster than human manpower. Also, it helps the farmers to increase the crop yield and reduce the operational costs [75].

The metaverse in Agriculture 5.0 provides numerous benefits, such as the creation of a virtual environment in which farmers can collaborate with trained professionals from around the world to share expertise and best practices. In Agriculture 5.0, the metaverse plays a vital role in training farmers to carry out harmful or dangerous tasks in a safe environment. For example, operating powerful machinery or handling chemical pesticides. By utilizing the metaverse, farmers in collaboration with machines can acquire the necessary skills and knowledge to reduce the risk of injuries and accidents [76]. The metaverse can simulate the complete development cycle of plants in a virtual environment which helps the users and robots to acquire information about plant growth in a quick time [77], [78]. The users and robots can learn the process of cultivating crops in a virtual farm environment starting from seedling to harvesting, so as to understand the various issues in crop cultivation. Users and robots can be trained in the metaverse by recreating a real working environment. Several smart farming helper applications have been developed by automating the entire smart farming process from cultivation to harvesting using AR, VR, and AI. The system consists of the automated application of fertilizers and water by detecting the movement of pests and soil moisture content. The system has various functionalities such as plant disease diagnosis, field monitoring, crop yield analysis, and monitoring of water stress and soil erosion. The metaverse will play a vital role in Agriculture 5.0 to increase food production and obtain maximum profit in the forthcoming years. Developing and implementing the metaverse for Agriculture 5.0 is complex and challenging. High-quality, portable metaverse models are required for implementation. Also, highly skilled training is required to make use of the metaverse as it is developed using VR and AR technologies and devices [79]. Fig. 4 depicts the application of the metaverse in agriculture 5.0.

## C. The Metaverse in Supply Chain Management 5.0

Supply chain management (SCM) helps industries to meet requirements and deliver customized products to consumers in a short span of time. The supply chain in the entire industrial process must be streamlined with current enterprise modeling for better adoption in the industrial revolution and other advancements [80]. Digital twin (DT) technology is applied to create a digital copy of the SCM processes such as inventories, warehouses, logistics, and assets. The DT helps in the complete lifespan of SCM from the procurement stage to the development stage and includes customer locations, suppliers, manufacturers, and factories. The simulation helps the DT to acquire the data from IoT devices which can be used by AI to forecast the hurdles in the various stages of SCM. This helps the manufacturers to take preventive steps to reduce the errors and losses in SCM. Collaborative robots (Cobots) also play a vital role in SCM processes. The dangerous tasks such as lifting heavy items, material assembling, transportation, packaging, checking the quality of products timely, and finally product delivery to the customers can be done by the cobots [7].

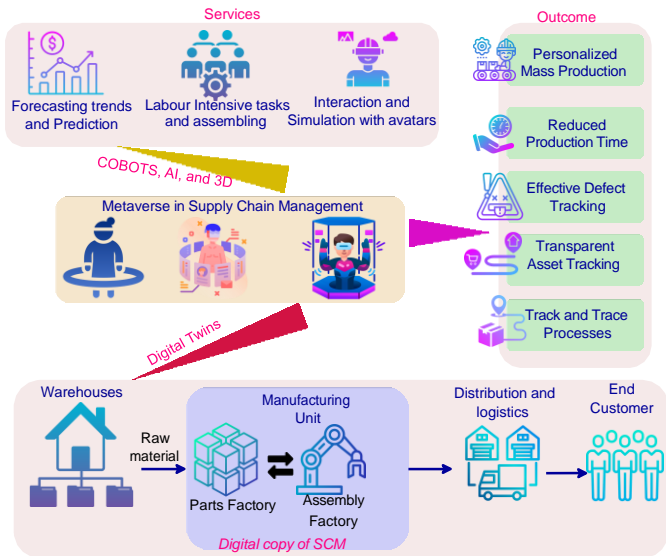

Fig. 5. The metaverse in Supply Chain Management 5.0

The metaverse can improve the effectiveness of the SCM starting from customer requirements to product development by enabling human and machine collaboration. Machinery, factory layouts, raw materials, and goods can be represented in virtual form using DTs. The developers can modify the shapes of the products, and analyze the various materials and goods by collaborating with experts all over the world in real time. This technology could help in reducing the production time. The manufacturers can share the 3D designs of various parts to identify the best material supplier. Warehouse workers use AR to trace the current position of products in the supply chain. VR is used to form realistic simulations of supply chain situations. For example, an expert driver in collaboration with VR trains warehouse staffs how to correctly use a fork truck. This would help in the inhibition of accidents and teach staff on how to use new tools. Machine learning algorithms are used to assess the day-to-day statistics of supply chain activities. Based on these statistics, the data analyst applies his knowledge and expertise to reduce costs, improve customer satisfaction, and make optimal decisions.

During the production process, virtual factories can be constructed using the metaverse to regulate the production plan and perform simulations in which machines and manpower can work together in manufacturing the products. Manufacturers can also construct virtual production lines to train the new employees through the metaverse [81]. A virtual warehouse can also be designed to have a stock layout plan before it can be transformed to a physical warehouse. The data in the physical warehouse can be integrated with the virtual zone so that the employees can find the number of goods, the location of the products, and the orders placed so far. Hence it is evident that the metaverse will enhance the SCM in the industrial sectors. The major challenges of the metaverse in the SCM are skilled labor, security, and privacy. Fig. 5 depicts the metaverse in SCM.

### D. The Metaverse in Healthcare 5.0

The advancements in the healthcare industry ensure humans' physical and mental well-being, which will promote a more significant contribution to a country's economic growth and industrialization [82]. Using AR/VR, telemedicine, AI, data analytics, and IoT technologies, the Metaverse Healthcare 5.0 system aims to provide effective and real-time medical services. These technologies, in collaboration with doctors and machines, form an interconnected healthcare system that improves patient care through remote monitoring, improves healthcare services, and strengthens healthcare research and education. The robots can diagnose and treat patients through proper guidance and instructions from doctors.

The upsurge of the metaverse has bought numerous prospects in almost all sectors. Some of the advantages of the metaverse in Healthcare 5.0 are as follows:

- Personalized treatments to the patients through wearables
- Social communal space among the healthcare community
- Medical IoT assisted with VR and AR for holographic construction with intelligent processing
- Exercises rehabilitation for elderly and disabled patients
- VR-aided therapies for nervous system diseases, anxiety and fear disorders, and post-traumatic stress

Yang et al. considered the metverse as the medical IoT aided with VR and AR glasses which are believed and expected to contribute widely to future computing platforms [83]. However, though the medical IoT has solved many problems like virtually connecting rural doctors with expert doctors worldwide, these expert doctors were not available all the time to assist the rural doctors. The expertise was also not available to supervise the research activities for clinical trials, and there is a lack of standards for real-time diagnosis through medical IoT. This paved the way for the metaverse in healthcare through medical IoT assisted with VR, and AR, that can overcome these limitations through holographic construction with intelligent processing to holographic emulation [84]. In turn, this will improve the control in virtual-real word interconnection, guaranteeing graded and customized treatment, thereby balancing the doctors' interaction on par with internal standards. Furthermore, the healthcare metaverse was suggested to be a good platform for exercise rehabilitation for elderly and disabled patients [85].

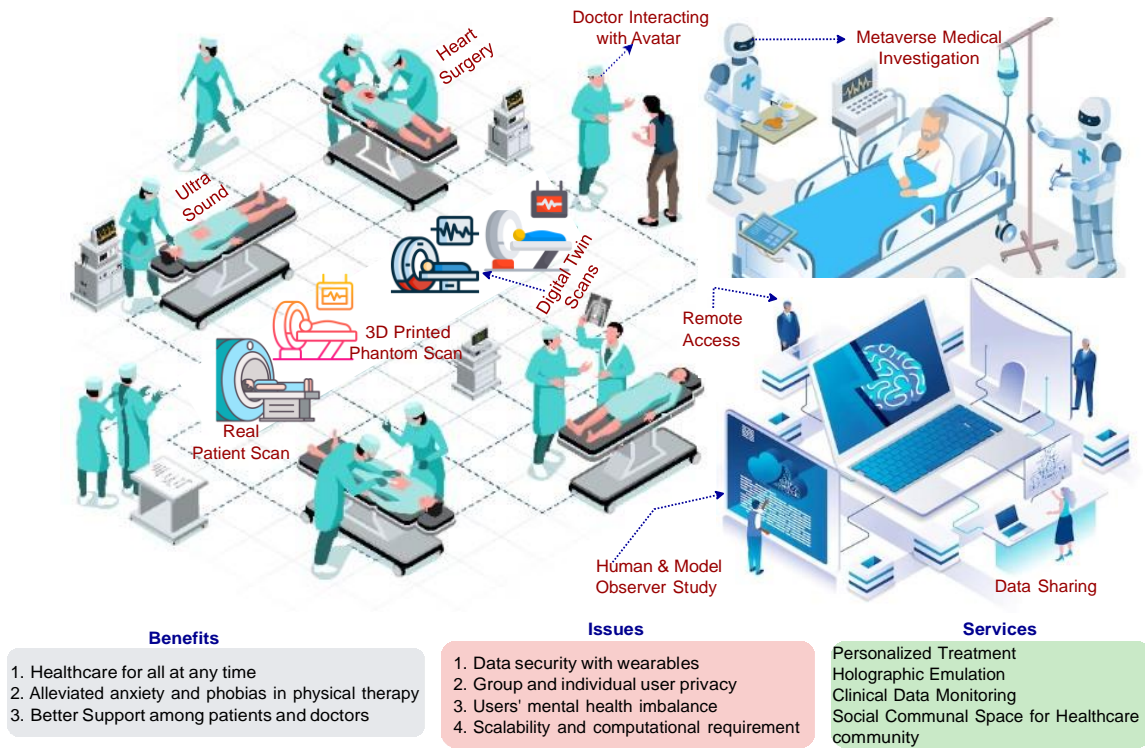

Fig. 6. The metaverse in Healthcare 5.0

Likewise, Liu et al. have highlighted the integration of various technological advancements for the benefit of healthcare stakeholders based on bibliometric analysis for healthcare metaverse. They have done bibliometric research on VR-aided therapy and investigated the major VR applications, namely, nervous system diseases, anxiety and fear disorders, post-traumatic stress, and pain management, to attain Health 4.0 [86]. Thus, VR-aided therapies can alleviate the physical therapy issues like anxiety and phobias. Also, VR-aided treatments can give better insights to healthcare experts into patients' emotions through the integration of physical therapy and psychotherapy, thereby creating a better rapport between the therapists and the patients. As a result, VR-aided therapies are efficient for the prevailing pandemic-like situation [86] and can be integrated with emerging advances to offer patient-centric customized therapy services. The metaverse has been employed in cardio vascular medicine where the doctor will meet the patient avatars for treatment and diagnosis [82]. Furthermore, Chen and Zhang have carried out a bibliometric analysis on the heath metaverse [87]. The health metaverse framework suggested will promote a better imprint on the medical field and social governance. As a result, the health metaverse was recommended for remote health monitoring, clinical data monitoring, orthopedic surgery, and patients' physical fitness through a 3D – immersive environment [88].

Although the metaverse-based healthcare provides efficient healthcare solutions, data security (unauthorized exposure of wearables), data privacy (group privacy and individual privacy), and users' mental health imbalance for the metaverse adoption are the challenges in its implementation. Another primary concern is scalability; when more doctors, patients, and other healthcare professionals simultaneously use the system, the response time and computational requirement will increase. Fig. 6 depicts the metaverse in healthcare.

### E. The Metaverse in Education 5.0

The metaverse has changed most of our daily interactions shift to the virtual environment which transformed the human lifestyle and community. The metaverse can help in various teaching and learning activities such as classrooms, museums, libraries etc. In this technology, students can navigate out of conventional classroom to a virtual environment [89]. It is very similar to video gaming which is very familiar to the students and it can propel them to come closer to the learning process in a very interactive, entertaining and enjoyable manner [90].

In the metaverse Education 5.0, simulations can be created in virtual world with the actions performed by avatars to expand the imagination and improve the collaborative intelligence. Teachers and students can immerse into the virtual learning environment. Students can explore the history lessons and the different parts of the vehicle using the AR gadgets. Using the gadgets, the students can move around and view the day-to-day happenings and the historical incidents through which they gain real time experience and knowledge. When mixed reality is integrated with VR technology, the students can experience the field trip in which they can physically communicate with the virtual ancient buildings and monuments. The metaverse education 5.0 provides personalized learning experiences for the students. Medical students can benefit from the metaverse in their education by visiting a virtual anatomy lab, performing surgery on virtual patients, communicating with virtual patients, and practicing clinical skills in collaboration with trained professionals. AI and machine learning techniques can further improve the quality of smart learning process in the metaverse environment. These technologies can be incorporated in digital avatars which can answer the questions asked by the students and give more alluring experience to them. Education metaverse can create virtual classrooms for variety of subjects such as science, English and history. Students can enter the virtual world protruded from their monitor using the gadgets. The teachers and students can easily transform to the virtual space because they simulate the original comfortable environment to them. Some of the application areas in smart education where the metaverse can be implemented are as follows: For learning subjects such as medical surgeries, aircraft driving, observing the internal parts of human body, group education visit to dangerous forests and travelling to archaeological locations [91]. The metaverse can further improve the smart education environment by integrating the various subjects into a unique virtual education platform for comprehensive learning experience.

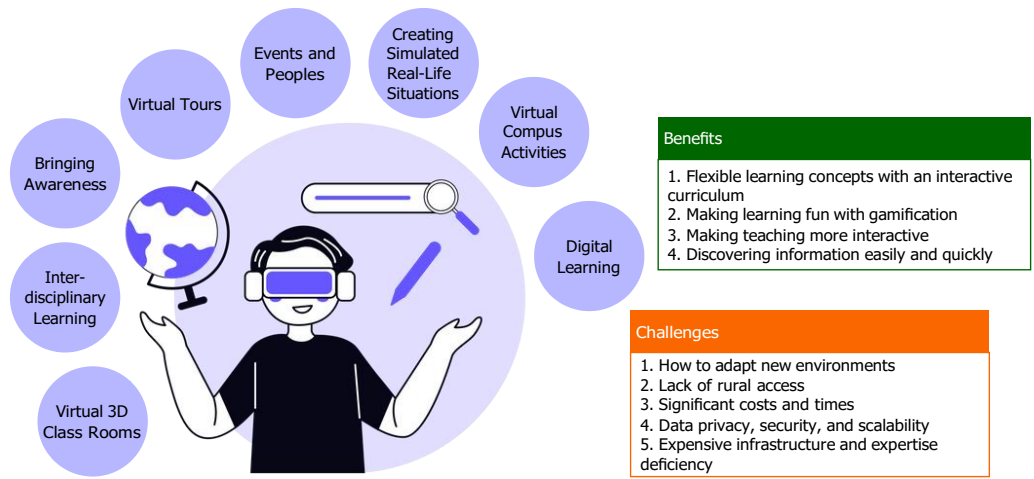

Fig. 7. The metaverse in Education 5.0

Some of the limitations of the metaverse in smart education are privacy violation of data collected and processed. The freedom in the metaverse may provoke the students to involve in unethical activities. The students in the metaverse immerse in the virtual environment which will cause confusion between physical and artificial world [92]. Fig. 7 depicts the metaverse in smart education.

### F. The Metaverse in Disaster Management 5.0

Natural disasters like floods, landslides, drought, storms, earthquakes, wildfire, extreme temperature, and volcanic eruptions may cause various hazardous consequences and human loss if not prepared properly. The metaverse for disaster management 5.0 provides an interactive and collaborative platform in which humans as well as robots can work together to react to disasters quickly and effectively. AR/VR simulations help in creating realistic disaster exercises wherein trained experts guide the public and staffs on how to react effectually to real disasters [93]. In actual disasters, this may assist them in enhancing their reaction speeds and decision-making abilities. VR applications like Second Life can incorporate real-life scenarios where avatars can be used to represent human actors and respond to events. In combination with trained humans, the metaverse can create a virtual environment where rescue workers from different organisations can collaborate and work together in a real-time crisis [94]. The AR-enabled metaverse can produce 3D maps by placing digital data on real-world images, which helps in estimating the level of destruction and finding survivors. For example, AR can assist in the creation of a 3D map of a destroyed building to help professional rescuers in their search for survivors. The AI-enabled metaverse analyses sensor and camera data to identify areas in a disaster zone that require human assistance. For example, AI can analyze the data from traffic cameras to identify traffic jams caused by waste, helping the government to focus on maintenance work.

Though VR-based simulations in collaboration with a skilled person can provide better disaster response training, it still lacks hands-on user experience and requires more skills for adapting to it. However, the metaverse world can overcome these limitations by enabling betterment in user experience [93]. The primary concern in the deployment of the metaverse for disaster management is scalability (certainly, a number of actor participants in this environment are unpredictable). In addition, human empathy for different deadly scenarios must be tackled with proper sensitivies. Fig. 8 depicts the metaverse in disaster management.

### G. The Metaverse in Transportation 5.0

The metaverse can connect the physical and virtual world enabling people to accomplish all their needs in one place without physical presence. Transportation in Industry 5.0 is the biggest vertical which requires enormous adaptation from the traditional way of transportation. The metaverse will change all facets of transportation such as aircrew transit, public tran-sit, logistics, staff transit, supply chain and intelligent transport systems. Fig. 9 depicts the metaverse in transportation 5.0.

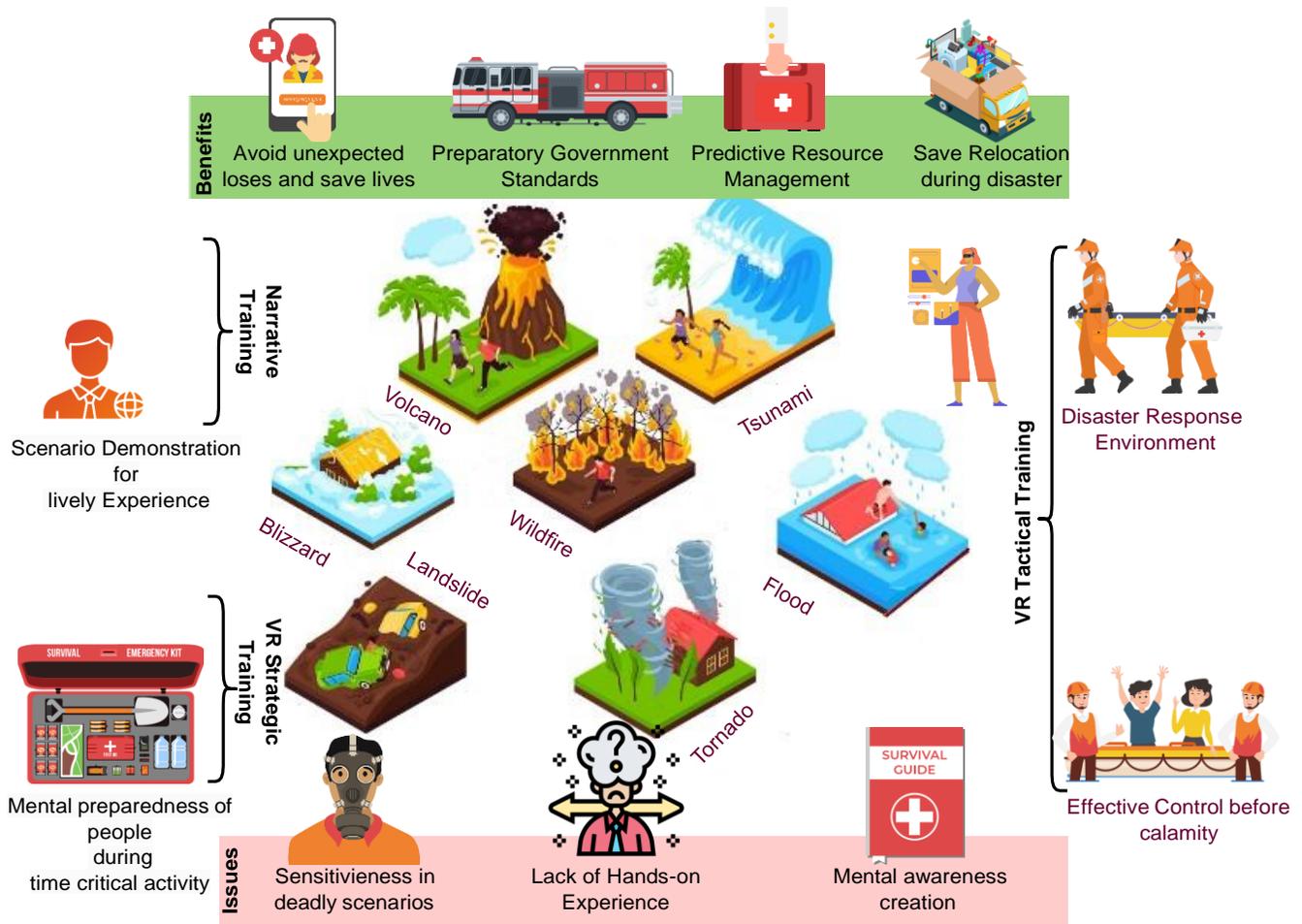

Fig. 8. The metaverse in Disaster Management 5.0

The metaverse can bring the following changes in the transportation:

- Robotic based mobility in inanimate objects and community spaces, named as Meta-Mobility [95]
- Leisure traveling by reduced travel time, cost, and energy
- Autonomous Vehicle Fault detection, repair and anti-theft systems
- Data-driven intelligent transportation
- Safer transport through advanced intelligence and governance

The primary concern in the metaverse is how transportation works. In the metaverse world, people can travel to different places without leaving their current location. In the metaverse, the transportation infrastructure changes as people will not use traditional forms of transportation, such as cars and airplanes. Instead, they use virtual transportation methods, such as virtual cars and virtual airplanes using AR/VR technology. AR/VR improves the passenger experience by entertaining passengers with games and videos. The integration of metaverse with Industry 5.0 allows passengers to choose their virtual vehicles, virtual environments, and virtual locations. Customization improves passenger happiness, comfort, and convenience by adding a personal touch. Virtual reality simulations can help nervous passengers reduce flight anxiety and make their trip more relaxing. Some of the enabling technologies in Industry 5.0, such as AI, IoT, and big data analytics, help in the advancement of smart transportation systems. These enabling technologies enable data processing, real-time data collection, and quick decision-making. The metaverse is a digital layer that integrates real-world transportation infrastructure, vehicles, and services in a virtual mode, allowing humans and machines to collaborate. In the metaverse, human-machine collaboration allows for human interaction and intelligent decision-making. The metaverse can be used as a training and simulation platform for autonomous vehicles, which improves transportation efficiency, safety, and user experiences. The collaboration of humans and machines has huge potential for shaping the future of transportation in the metaverse.

The metaverse will have an impact on employment and layoffs in Transportation 5.0, particularly for blue-collar employees. The metaverse may lead to changes in employment duties and specifications in Transportation 5.0. For example, autonomous vehicles replace human drivers, decreasing the demand for drivers. It also provides a wide range of op-portunities for technical professionals during the development and maintenance of metaverse-enabled transportation. With the help of advanced technologies and specialised skills, technical experts contribute to the development of transportation systems and play an important role in the successful integration of the metaverse in Transportation 5.0.

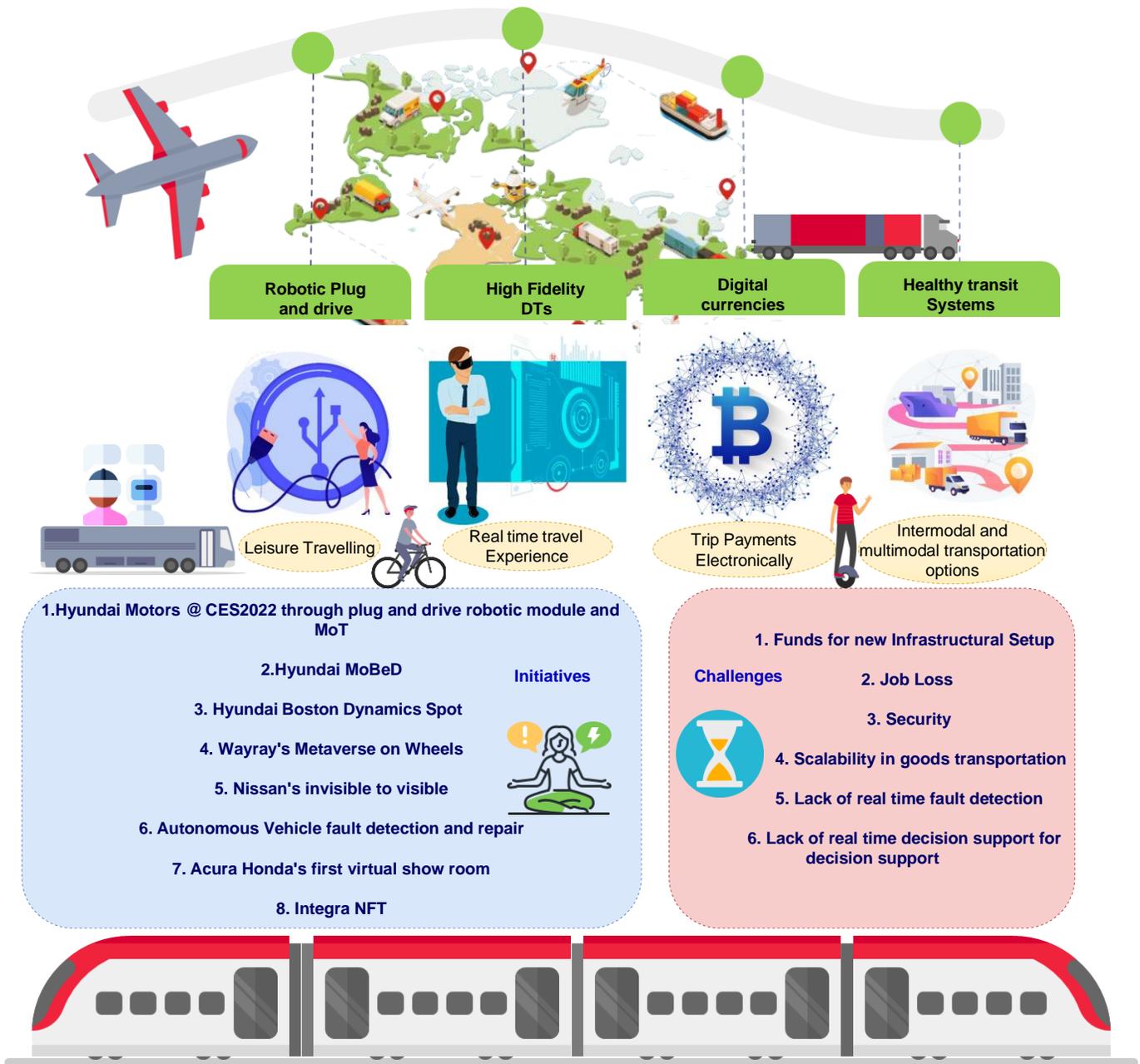

Fig. 9. The metaverse in Transportation 5.0

Hyundai motors has released a press note at CES 2022 stating its future vision on "Expanding Human Reach" through robotics and the metaverse to realize humanity's unconstrained liberty of movement [96]. They have also revealed the concept of Mobility of Things (MoT), enabling mobility in inanimate objects and community spaces through robotics (Plug and Drive module). Mobile Eccentric Droid (MobED) and Boston Dynamics' Spot are some of the company's robotic products which can carry anything virtually towards attaining its brand vision of progress for mobility. Furthermore, metaverse can play a vital role in intelligent transportation systems [97]. They have used two case studies namely Wayray's Metaverse on Wheels (holographic deep reality display in cars) and Nissan's invisible to visible, and discussed various challenges in different aspects of intelligent transportation.

For the metaverse to replace today's transportation infrastructure, the economic losses and funds for a new infrastructural setup must be realized. The numerous job loss to people in the transportation sector must be replaced with equivalent job opportunities. In addition, the implementation constraints like security and scalability should be accounted for during the transit of physical goods.

### H. The metaverse in Smart City 5.0

The smart city is the digitization of modern cities through various Information and Communication Technologies (ICT). Everything is interconnected to share data to make effective decisions in improving the citizens' welfare and effectively imparting government policies. Smart city 5.0 aims at sustainable city management by utilizing modern technologies where human and artificially intelligent agents (robots) will be working collaboratively for optimized city life management to balance all spheres of different city actors harmoniously [98]–[100]. Smart cities collates the information from the various sources like cameras, sensors, social media for regular feedback services to the policymakers for the betterment of the services [5], [101]. Smart city 5.0 strives to ensure urban resilience for optimal utilization of resources [102]. Fig. 10 depicts the metaverse in smart city 5.0. The integration of the metaverse in Smart City 5.0 results in a modern urban environment that improves people's quality of life while also promoting environmental sustainability. The metaverse in smart city 5.0 enables citizens, businesses, and government bodies to virtually collaborate, communicate, and share information in collaboration with trained professionals. The integration of the metaverse and digital twins in Smart City 5.0 helps the architects to construct virtual prototypes, model designs, and improve plans before executing them in real time. Human collaboration enables the creation of effective and environmentally friendly city layouts, allowing executives to take efficient decisions and improve the overall urban environment. The metaverse provides virtual tours that let travelers from all over the world to sightsee the city's important places. The metaverse helps in monitoring city traffic, water waste, electricity, and public safety in smart city 5.0 [103]. To encourage tourism, the South Korean government launched a smart tourism city project in Incheon. The project's goal is to use the metaverse to improve people's tourism experiences, giving visitors a more pleasant and personalised experience [104]. Incheon Easy is a smart tourism application offering two metaverse services. ARIncheon provides a real-time tourist experience using AR via the smartphone camera sensor and IncheonCraft, which integrates Minecraft (a sandbox game) to give the tourists an Incheon experience. IncheonCraft is a virtual-based metaverse where the tourists can experience the tour as avatars. At the same time, ARIncheon is a real-based metaverse by providing the historical maps in digital display and offers operational Incheon incentive-based service for engaging the tourists.

Lim et al. [105] have presented a case study on the metaverse cities like "Metaverse Seoul" with a collaborative edge-based framework for virtual city development in the metaverse. They have recommended that edge intelligence can be leveraged to attain the features of the metaverse with its low-latency communication and faster response [105]. The various products of meta for the smart city include Horizon Home (meta's social media platform), AR calls( holographic and video calls), Virtual gyms, Presence Platform (Meta development kit), Project Cambria (Virtual Headset), and Spark AR ( for meta creators community) [106].

The data privacy and security in sharing users' private data (such as civil services) must be strengthened. Also while experiencing real-time experience, the physical uncertainties must be mapped for a better experience.

### I. The Metaverse in Cloud Manufacturing

The cloud manufacturing (CMfg) is one of the most demanding networked manufacturing paradigms. CMfg is an integration of ICT with advanced manufacturing techniques throughout the manufacturing life cycle processes. CMfg is one of the key technologies of Industry 4.0, enabling the manufacturing with a single click of the user requirements. CMfg has a manufacturing cloud, resource layer, virtual service layer , application layer, interface layer and global service layer. The manufacturing cloud encompasses the different manufacturing service providers, their resources, and services to meet the needs of different clients on a demand basis. This export process in CMfg realizes a multi-agent collaborative environment in the manufacturing resource layer where different manufacturers can share their resources such as 3d printers, robotic arms, machines, design tools, simulation tools, and modeling software. Also, it enables sharing of the manufacturers' capabilities in rendering different services. The core support ensured by the CMfg will assure efficient resource management and effective search of resource needs in addition to manufacturing life cycle services encapsulation, through the virtual service layer. The dedicated collaborative applications are provided by the application layer of CMfg. The interface layer helps the consumers to seek different services like production as a service, experimentation as a service, design as a service, simulation as a service, manufacturing as a service, maintenance as a service, and integration as a service. The global service layer maps the consumer requirements with the multi-agent collaborative environment to offer a predictive and customized user experience. This enables the user to monitor their product throughout its development life cycle. All these are possible through integration technologies like IIoT, DTs, CPS, AI and wireless sensor networks. Some of the common issues reported by the practitioners and reporters of CMfg are variability in production planning with uncertain user demands, vulnerability to security threats, social acceptance, faster multi-agent collaboration, regulation compliance, and huge maintenance overhead [107].

The metaverse in the cloud manufacturing can assist in the following ways:

- Individual avatars can design and test the customized component in the immersive environment
- Multi-stakeholder social communal spaces such as end-user, designers, manufacturers, and operations teams can discuss and collaborate through their avatars for operational efficiency [108]
- Faster and simultaneous design and production
- Forecasting inventory using on-demand sensing AI models
- Improves scalability in uncertain user demand using predictive analytics thereby reducing lead time
- Low-latency response in the data processing through edge enabled metaverse platform during larger user demands

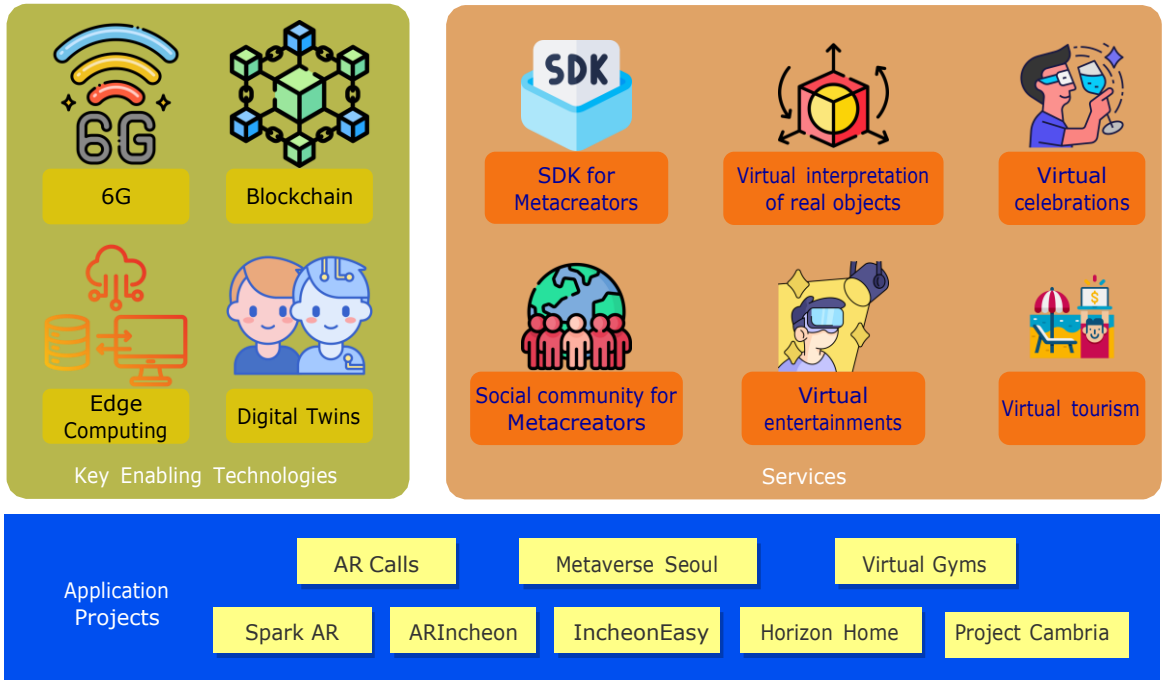

Fig. 10. The metaverse in Smart City 5.0

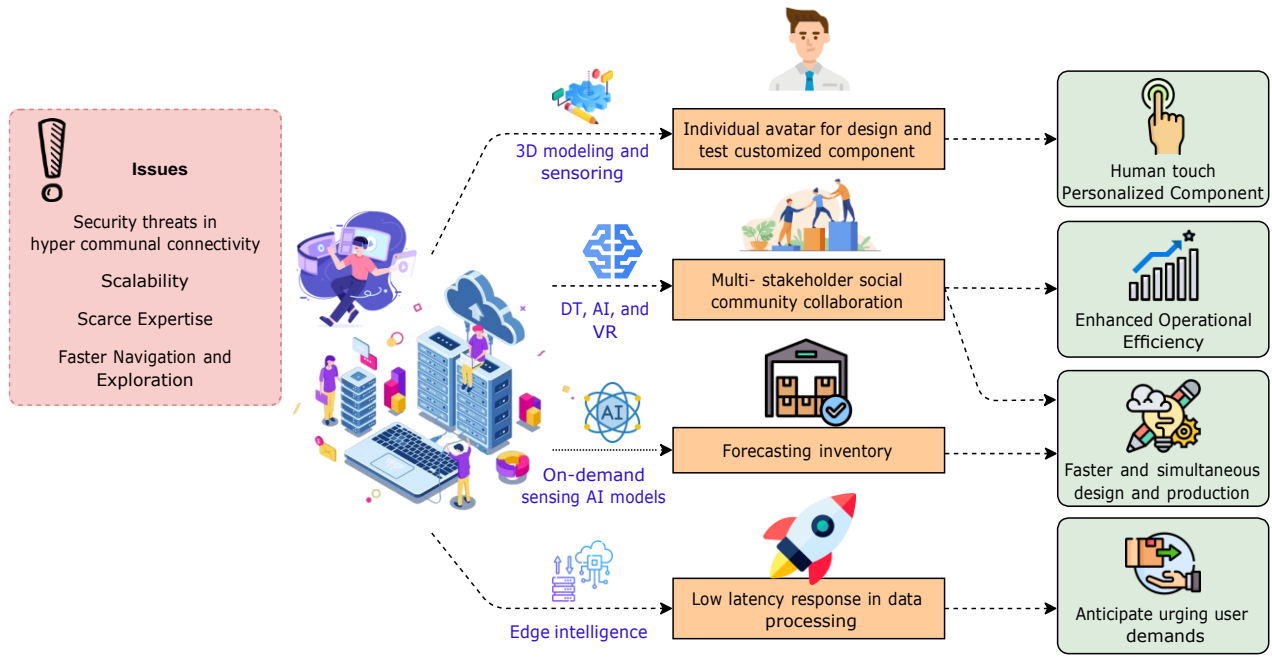

Fig. 11. The metaverse in Cloud Manufacturing

- Digital twins in the metaverse can reduce the quality control risks

The major issue in incorporating the metaverse with cloud manufacturing would be security threats in hyper communal connectivity, and interoperability among heterogeneous service providers [109]. Furthermore, scalability in terms of CMfg consumers, as well as the general metaverse community users, will be another hectic issue. Fig. 11 depicts the metaverse in cloud manufacturing.

### J. The Metaverse in Robotic Automation

As Industry 5.0 aims to bring the human touch back to the production floor and development by teaming up humans and collaborative robots, robotic automation is one of the most important components of Industry 5.0. Though the cobots depend on human intelligence, they must be automated in such a way to cope with human commands without any consequences in the new production environment. Here the cobots are intended to do labor-intensive tasks and to ensure consistency with skilled humans' cognitive skills [10]. The metaverse utilizes robotic automation more effectively because of its smart functionalities. In the metaverse environment, the human avatars (the human virtual image) and cobots will be collaborating to carry out the production tasks. Cobots must be automated and trained to interact with human avatars as well the real human. The metaverse in robotic automation has the following:

- Cobot and Avatar collaborate to make the metaverse life reality on the production floor
- The metaverse supports the creation of 3D models of the patient's anatomy, while cobots assist doctors in carrying out an operation and improve patient safety, operation time, and patient satisfaction
- A cobot in collaboration with the metaverse creates a virtual training environment where staff members can practice new skills with no risk of injury.
- The metaverse helps in remote maintenance and repairs by allowing professionals to remotely access machines, examine them, and then carry out virtual repairs.

The real-time motion mapping of avatars or robots with that of humans is challenging. Spanlang et al. have suggested a technique for controlling the real-time motion streaming of avatars and robots. A case study on the embodiment of a person as an avatar in an immersive environment was presented to demonstrate the effective remapping of joint positions, bone length, and rotations of the human skeleton and its avatar or robots [110]. Hyundai Motors has disclosed its decision on "meta-mobility" through robotics automation in the metaverse. China has introduced its first metaverse robot through the integration of IoT, AR, VR, AI ad other cutting-edge technologies. The robotic skin by Meta was specially designed using thin rubbery plastic with magnetic particles to feel the sense of touch. The special robotic skin, with less than 3mm with AI was developed in collaboration of Meta (Facebook) and Carnegie Mellon University [111]. Unity platform by San Francisco has used robotics to train and design the metaverse world [112]. Also, NVIDIA Omniverse and Microsoft have used robotics for connecting two worlds (real and virtual) within the metaverse using 3D assets and the creation of enterprise metaverse respectively [3]. The major challenges in robotic automation are vague use cases, continual upgradation and maintenance, improper governance, lack of standard automation processes, attaining convincing expectations, and skilled employees. Fig. 12 depicts the metaverse in robotic automation.

The various technical requirements for different verticals of Industry 5.0 and its coverage in the applications discussed in this section are summarized in the Table III. The various technical requirements in Table III are:

1) TechR 1 ← Computation Power
2) TechR 2 ← Memory Management
3) TechR 3 ← Scalability
4) TechR 4 ← Accessibility
5) TechR 5 ← Interoperability
6) TechR 6 ← Security and Privacy Issues
7) TechR 7 ← Legal Issues
8) TechR 8 ← Skilled Professionals
9) TechR 9 ← Brain-computer Interfaces
10) TechR 10←Inter dependencies among different societal applications

The technical requirements include interdependencies among socio-economic applications as all the smart applications or different verticals in industry 5.0 will work collaboratively. Therefore, rather than interoperability among devices, inter-dependencies among these applications matter a lot in an industrial metaverse environment. In addition to generic interfaces used in applications today, the metaverse environment requires interfaces assisting the gaming environment that promotes the connection between brain and computer, referred to as brain-machine interfaces [113].

## IV. THE METAVERSE IN INDUSTRY 5.0: RESEARCH PROJECTS

This section highlights some of the key research projects related to the metaverse and Industry 5.0.

### A. WayRay

WayRay is a Transportation 5.0 based project that usages the metaverse to advance automotive transportation. Their AR heads-up display technology has the potential to make transportation more efficient, secure, and eco-friendly. Their system helps drivers stay informed and alert by projecting crucial information onto their windscreens, such as directions and traffic conditions. Moreover, their AR technology allows the production of fully immersive experiences for travellers by supplying appropriate context data. WayRay's transportation technology helps drivers safe with predictive warnings and alerts, keeps them productive with dynamic traffic updates, and tracks their fuel efficiency and vehicle emissions. WayRay has partnered with major manufacturers and is taking steps to get its technology to market. With the metaverse, WayRay can make automobiles more secure and efficient, bringing in a new era in transportation [116].

### B. Nikeland

Nikeland is a project that was developed by Nike and integrates the metaverse with SCM 5.0. It provides a virtual environment for its users to shop for Nike merchandise, play games, and interact with other users. Nikeland provides the accountability and traceability of virtual objects by using features of blockchain technology. This allows users to virtually try on different varieties of shoes using AR technology, which enhances the user's buying experience. Nike will also be significantly benefited by acquiring data insights about their product design and manufacturing process, which will help them improve their marketing strategies. Nikeland transformed traditional retail and SCM by providing an engaging and data-driven environment [117].

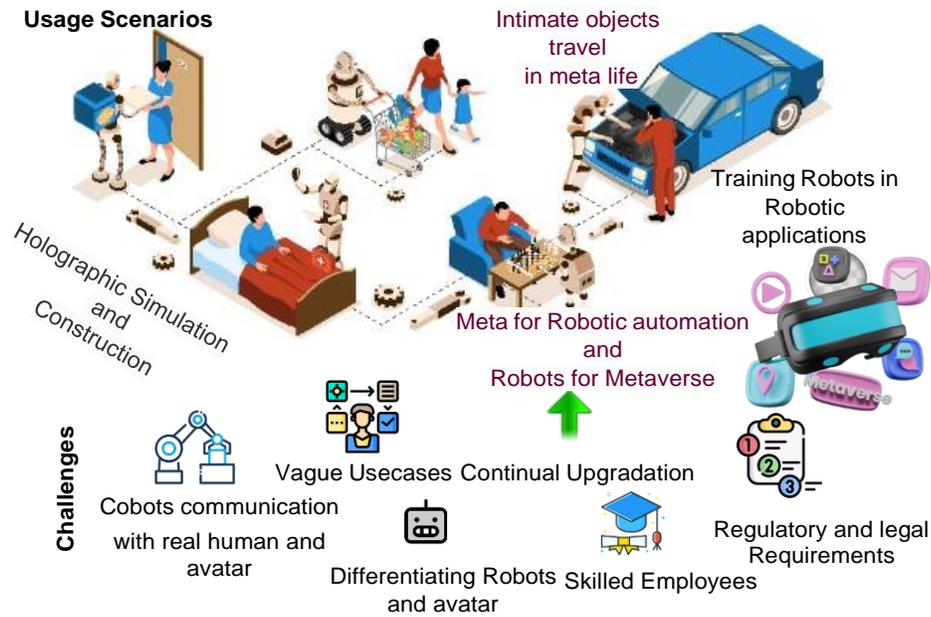

Fig. 12. The metaverse in Robotic Automation

TABLE III
TECHNICAL REQUIREMENTS FOR DIFFERENT VERTICALS OF INDUSTRY 5.0.

| S.No. | Verticals of Industry 5.0 | TechR 1 | TechR 2 | TechR 3 | TechR 4 | TechR 5 | TechR 6 | TechR 7 | TechR 8 | TechR 9 | TechR 10 | Remarks |
|---|---|---|---|---|---|---|---|---|---|---|---|---|
| [1] | Society 5.0 [66], [71], [72], [74] | M | M | L | H | N | H | L | L | M | L | Focussed on the implementation perspective of a metaverse in almost all smart applications of the society. |
| [2] | Agriculture 5.0 [75], [77]–[79] | M | M | M | L | M | H | L | L | L | N | Covered the few technical perspectives but ineffective discussions on regulatory issues in technology adoption and accessibility. |
| [3] | Supply chain Management [81], [114] | M | M | L | M | L | M | N | M | M | L | Though the virtualization of the entire supply chain was covered, interoperability concerns and legal opinions were less discussed. |
| [4] | Healthcare 5.0 [82], [83], [85], [86] | M | N | L | L | L | L | N | L | L | M | Focused on the adoption of a metaverse in almost all perspectives of healthcare, but memory constraints and legal issues with data exposure are merely deliberated. |
| [5] | Smart Education 5.0 [90]–[92] | N | M | M | L | L | M | H | M | M | N | Broader perspectives of a metaverse in educational institutions were focused but practical implementation constraints are not explored. |
| [6] | Disaster Management 5.0 [93], [94] | N | L | M | L | N | L | L | L | M | N | Proper sensitiveness must be assured while handling deadly scenarios which impact human empathy. As well, scalability must be focused on in a deeper sense. |
| [7] | Transportation 5.0 [96], [97] | M | N | M | M | M | H | N | M | L | N | The economic losses and funds for a new infrastructural setup must be realized. Security and Scalability must be accounted for physical goods transit. |
| [8] | Smart City 5.0 [98], [103]–[106] | L | L | L | H | H | H | H | L | M | L | Data sensitivity, privacy, and security are mandated requirements when dealing with public services. Security breaches with apps in the gaming world must be assured. |
| [9] | Cloud Manufacturing 5.0 [107]–[109], [115] | M | L | M | M | M | H | M | M | N | N | Memory constraints, and associated storage security threats in this hyper communal space require more attention. |
| [10] | Robotic Automation [3], [10], [110]–[112] | L | L | M | N | L | M | L | M | M | N | Robots use metaverse and metaverse use robots. Therefore, major discrimination between avatars and robots must be detailed with proper training requirements. |

| H | High Coverage | M | Medium coverage | L | Low Coverage | N | Not Discussed |
|---|---|---|---|---|---|---|---|

## C. Minecraft Education Edition

Minecraft Education Edition was developed by Xbox Game Studios and Mojang Studios mainly for use in educational environments. Minecraft Education Edition is an application that integrates the metaverse and Education 5.0. It offers students an immersive learning environment in which they can explore topics using virtual environments and interactive activities. The platform encourages collaborative learning, personalization, and global access for professionals. Minecraft Education Edition transforms education by encouraging creativity, engaging students, and preparing them for the digital era by using cutting-edge technology and gamification [118].

## D. Landian

Landian is an metaverse based agriculture 5.0 project, which allows purchasing of land and digital assets in the metaverse. It provides a platform for farmers to interact, share knowledge, and teach new framing techniques to each other. Landia uses NEAR protocol which is based on binance blockchain specifically designed for the metaverse platform. Landin provides farmers various benefits which includes the ability to collaborate and share farming methods virtually. Farmers can also gather relevant educational resources related to farming and can exhibit their agricultural products in the virtual marketplaces. These marketplaces also allows farmers to buy and sell agricultural products. Landian shows a huge potential to transform the agricultural sector by integrating the metaverse [119].

### E. HealthBlocks

HealthBlocks is a startup with an aim to inequality in global healthcare. It is a project which integrates metaverse with Healthcare 5.0. It uses blockchain technology to provide decentralised and secured access to healthcare services. The users can communicate with doctors from anywhere, access medical information, and obtain doctor's prescriptions in a virtual environment provided by the metaverse. HealthBlocks will help in increasing the effectiveness and quality of patient treatment by making healthcare available and budget friendly to everyone. Users will highly benefited by its simple access to healthcare services from anywhere around the world. HealthBlocks looks like a promising solution that has the huge potential to revolutionize healthcare sector with the help of the metaverse [120].

## V. CHALLENGES AND FUTURE DIRECTIONS

In this section, we highlighted the integration challenges of the metaverse for Industry 5.0 along with possible future directions.

### A. Research Challenges

Automation in everything will almost raise issues like higher computational requirements, data security, user privacy, health issues, and joblessness from the users' perspective. But, on the other hand, from the developers' and technology seekers' perspectives, providing a more dynamic and responsive human-machine interaction, lack of proper standards, security vulnerabilities, and the adoption of the technological advancements will be the more imperative challenges. This section highlights the research challenges, feasible solutions, benefits, and further directions for future research in implementing the metaverse. Challenges in the metaverse ecosystem and possible benefits in mitigating those challenges are shown in Table IV.

*1) User Interaction:* The metaverse employs several interactive devices to immerse the users into a virtual world. These devices are being developed to full-fill the user requirements. The features of user interactive devices which will help to interact with the virtual world should be portable, lightweight, comfortable and wearable. The medium of interaction should be transparent in such a way that the technology should be invisible to the users and they should get drown themselves into the virtual environment. Some of the interactive technologies and devices are VR, AR and MR. VR is a virtual environment created artificially and users can immerse in the same way like they do in the real world. AR is a technology through which users interact with virtual world and it depends on wearable devices which helps to merge digital objects with live video. MR is an advanced technology in which the physical environment performs interactions using the collected digital data. Most of these interactive devices are portable, lightweight and comfortable. The high expensive cost of the interactive devices is one of the major challenge in adopting the virtual technology. Prolonged use of VR headsets will create psychological issues, neck and head tiredness. Apart from these hardware challenges, there are some issues in the quality of developed the metaverse models. Therefore high quality VR, AR and MR gadgets and reliable models can help the users for better interaction [91].

These issues can be avoided through realistic effects in a visual rendering, enabling users to touch and feel the effect like ball bounces or water rippling, but tactical emotions are still difficult [128]. To provide a sustainable interface, holograms, and eye lenses can be used alongside head-mounted devices. Furthermore, they have suggested that in addition to the human players in the virtual world, non-player characters' interaction with the users must be considered. To capture and learn about different sensory actions of humans, multimodal pretrained learning models can be utilized to learn more about visual language movements [129]. This can help in avoiding the misinterpretation of human emotions by avatars.

The integration of the metaverse with Industry 5.0 is challenging since decisions must be made by both humans and machines. Due to the fact that humans cannot directly touch or feel the physical prototype provided by machine, they may be misled while making judgements. Furthermore, the interaction between humans and machines must be uninterrupted, and network or bandwidth delay will lead to catastrophic situations. The devices or equipment now used for interaction may not be suitable in all verticals of Industry 5.0 and need customization based on the applications requirements. Therefore, creating a perfect virtual replicate of a real industrial machinery is still a challenge to be addressed.

*2) Computing Resources:* One of the most prominent challenges in implementing the metaverse is the convergence of various heterogeneous technologies like IoT, VR, AR, AI, gaming, and so on by accommodating varying computational requirements. Also, colossal data accumulation may require higher and faster computing power with ultra-low latency.

The metaverse is like a massively multiplayer gaming environment that allows multiple users to participate simultaneously, requiring higher GPU and HMDs to render an immersive virtual world balanced with the physical world. So, edge computing with cognitive edge services will be an essential candidate for the metaverse to serve ubiquitously. By leveraging the AI, local decisions like positioning an avatar can be made at end devices (like the engine).

Furthermore, the expensive foreground rendering (which requires lower latency and fewer graphics) can be done at edge servers instead of the cloud servers to reduce end-to-end data latency [121], [130]. The aggregated data can later be communicated to cloud servers. At the same time, the higher computational intensive background rendering can be done at the cloud servers. Consequently, AI models can be utilized for resource allocation in these denser user distributions to improve the Quality of Experience. Therefore, AI-enabled edge computing would help to solve the computation overhead issue [49].

To guarantee, an immersive experience in the virtual world as that of the physical world, the metaverse imposes a wider set of requirements like reduced motion-to-photon latency, real-time rendering and control, and high-quality visual appearances. Mobile edge computing empowered with 6G will be an essential candidate to serve this requirement [131]. The latency requirements, network load, and computational resource allocation can be effectively managed through the convergence of the metaverse with mobile edge computing, blockchain, 6G technologies, and AI [132]. Mobile edge computing will be an essential part of any grander technological amalgamation for deployment on a larger scale.

TABLE IV
Challenges in the metaverse ecosystem and possible benefits in mitigating those challenges.

| Challenges | Solution | Technology | Benefits |
|---|---|---|---|
| Higher Computational and storage requirement [121] | The metaverse framework with AI-enabled edge computing will require higher computational resources, communication requirements and huge storage | Meta blockchain with cognitive edge | Ultra-low latency response. |
| Centralized Virtual goods trading [122] | NFTs can be used for trading the virtual items for replicating their use across the metaverse | Meta blockchain with 6G networks | Secured and decentralized virtual goods trading |
| Data Interoperability [109] and Seamless traversal among user data at different virtual worlds [30] | blockchain with cross chains can store the data in a structured format and helps the two different blockchains to interact | Meta blockchain with cognitive edge and 6G networks | Digital safety through secured data interoperability |
| Human Workload requirement on HCIs | The combination of VR and AR may reduce the users' workload requirements [123] | Prolonged Usage testing using a Regressive testing framework | Users may feel it easy to adapt to the technology-mediated environment |
| Content integrity and user privacy [124] | Data scavenging to avoid unauthorized data access and pollute the environment | Meta blockchain with Federated learning and 6G networks | Safe and privacy-preserving content delivery |
| Copyrights for digital assets [125] | Purchasing virtual items through NFT requires copyrights permission for the creators and verification of the same | blockchain with smart contracts and quantum computing | Ensures asset confidentiality and integrity in digital purchase |
| Portable and comfortable user Interactive devices [126] | User interactive devices are not portable, lightweight and comfortable. The cost of these devices are expensive. Also long use of VR headsets leads to pain in neck and head | Prolonged Usage testing using a Regressive testing framework | Portability and transparency to immerse the users into digital environment |
| Open platform with new standards and principles for developing the virtual technology [127] | The metaverse environment controls the emotions and behaviour of the users which may lead to destructive disasters. The user obsession to this digital environment will become unavoidable in future | Adopting standards for VR adoption | Psychological issues can be minimized |
| Moral and ethical standards to publish the content in the metaverse [4] | Users can be provoked to involve in criminal activities such as robbery, sexual abuse and assaulting. Publishing wrong information to the users lead to exploitation and misconceptions | Adopting standards on sensitive data | Controls misconducts and criminal activities. Provides safe user environment |

Thus, the integration of the metaverse with Industry 5.0 requires massive processing power to operate all of the enabling technologies and devices, which is significant challenge to be addressed. To address the computational needs of the metaverse across multiple Industry 5.0 sectors, high performance computing systems and massive cloud infrastructures are also required. A network with strong data transmission capabilities is also required for efficient usage of the computational systems.

*3) Security and Privacy:* In the social virtual world in today's plurality of the Internet with heterogeneous sources, the users find the communities matching their preferences, thus preferring to have a singleton match for their choices. This requirement towards the search for a singleton online community of preferred choices is one of the main reasons for the emergence of the metaverse. On the other hand, the uniqueness of the metaverse is a singleton by aggregating various technologies, services, and goods as a shared and centralized service point. Furthermore, security in the metaverse comes with two different verticals: data security and software security. So, this centralized metaverse may force the new users (peculiar interest) to participate in its environment, and many behave weirdly to compete with other users (some may exploit too). AI singularity can make AIs in the metaverse life unaware of itself and continuously improve to perform better [122].

The convergence of cyber-physical worlds with exponential technology growth has led to open security and privacy issues. Also, the highly immersive, interconnected, and interoperable environment may allow the participants to trade the virtual items online as Non-fungible tokens (NFT). These virtual items can be used in all other spaces of the metaverse [122]. This may even create unprecedented security vulnerabilities. Most people today rely on online shopping, leaving the footprint of their desires and helping the social network analyze and predict the users' needs, thereby being the product of the Internet. In turn, these users and everything will be the product of the metaverse, and the meta platform will provide surplus information about all the users to the content creators and similar businesses. As a result, users' privacy is in jeopardy in unsuspected ways. Like at a particular instance, the metaverse can closely monitor our body movements, brain responses and can predict where the user can click, on what items, and how much time they will spend [133]. The users' personal information, behavior (physiological characteristics), and interactions are three perspectives of user privacy in the metaverse. Through doxing, the users' data are already misused for online shaming [134]. Doxing in this strong bonding physical-virtual world will offer exceptional opportunities for hackers to exploit the virtual world's immersive experiences and harm in the physical world. Like other social engineering attacks, stalking and spying [135] in the virtual environment will be experienced more in the metaverse. Other notable attacks are cyberbullying, shit storming, video call bombing, gender harassment (through sexting), and raiding. These forms of denial of service may ruin the participants in the metaverse.

Data integrity and user authentication are the most critical security issues to be cited while relying on automation and algorithms to maintain a virtual world. Many fabricated contents and unauthenticated users will be replicated more in the metaverse. It has been envisaged that software-driven accounts can be fingerprinted or digitally reproduced in the social media network. Furthermore, advanced AI models can make more automated accounts without noticing or detecting any algorithms. This will raise more fatal security threats in the immersive physical – virtual world environment. Blockchain based solutions can be incorporated to solve these security, integrity and user authentication issues [124]. Kwon et al. suggests that Quantum-based metaverse ensures faster and more secured metaverse applications through their proposed case scenario MetaQ which utilizes quantum kernel ML [136].

As in any technological advancements, Security and Privacy remains a significant challenge upon the integration of the metaverse with Industry 5.0. The metaverse enables large user engagement across different Industry 5.0 sectors, and users have to share sensitive information with these applications to access the services. Any attack on such sensitive data could compromise user data privacy, which is a problem that must be addressed. One of the fundamental elements of the metaverse is user anonymity, which raises concerns regarding accountability and transparency, potentially affecting the concept of Industry 5.0. This integration also opens up the new possibility to the creation of new virtual assets depending on the Industry 5.0 sectors and, in turn securing them will be further challenge that must be addressed.

*4) CyberSyndrome: Social Media Addiction and CyberBullying:* There are many emotional challenges faced by the metaverse due to the fundamental technologies such as VR and AR on which they are being developed. These technologies control the behaviors, emotions and intelligence of the users. The diversion of users attention in locality based metaverse applications can end up with destructive disasters. The most common health issues related to the virtual technologies are drowsiness, blurred-vision, nausea and motion syndrome. In a recent survey, nearly 50% of adults utilize social media in their day today life during pandemic period. Due to this advanced virtual technology, the user obsession to the metaverse is inevitable in future [137]. Users may depend on these virtual environment to elude from the physical world. The situation may become even dangerous where AR and VR can redirect the users to dangerous activities such as burglary, assaulting and other criminal events. The other issues the metaverse developers need to address are mental disorder issues, sexual harassment and other exploitations in virtual space such as sexual abuse and harrassment. Unethical behavior may be more dangerous in the metaverse than in physical world. VR can immerse the users in virtual environment where sexual abuse in the imaginary world can be sensed as real experience. Sometimes the users may face time syndrome where they can not able to distinguish between the timings of physical and virtual world. When the children learn through the metaverse, there is also possibility that some intruders can provide false information to exploit them and engrave misconceptions. The enterprises who develop the metaverse platforms should apply strict security solutions to ensure that the virtual environment will not be attacked by the external hackers and internal attackers [4]. A keypad application named catch a word program was developed to safeguard young Indonesians from cyberbullying type of issues in social media [138]. This can be installed in users' smart devices to avoid rudeness in digital media communication. Shielding can be offered through the words or text rendered by the avatars on the scene. But predicting the behavior of an avatar is difficult as avatars' behavior is diverse and delicate. The detection of malicious activity (bullying) by an avatar can be achieved by considering multiple factors, such as gestures, facial recognition, emotion, and social interaction [139].

The incorporation of the metaverse into Industry 5.0 presents consequential challenges in terms of work addiction. The metaverse provides a platform for experts or humans to interact with machines that is always accessible and available. Because of this seamless integration, individuals will find it difficult for stabilizing their personal and professional lives, which in consequent may lead to mental imbalance and may prone to health disorders. Therefore, better means for stabilizing these concerns upon industrial metaverse implementation must be addressed. The integration of the metaverse with Industry 5.0 will result in a highly competitive work culture, which will place enormous strain on employees and may lead to stress and health problems.

*5) Impact of Human-Computer Interfaces(HCI) on Human Mental Workload:* The advancement of technologies and HCIs with higher computing power has led our lives to a technology-based ecosystem. The metaverse uses VR, AR, XR, or MR technologies to connect the physical world with the virtual world. The major challenge is dropping the human workload requirements while participating in the metaverse ecosystem, and in analyzing the metaverse environment's usability based on user involvement. The human workload can be determined by weighted perceptions of different factors such as physical demand, mental demand, performance, effort, temporal demand, and frustration [123]. The effect of XR on six dimensions of human workload has been investigated in [140]. They observed that AR significantly impacts the mental demand and effort dimensions of the human workload. On the other hand, VR doesn't affect human workload as the VR interfaces are very natural as current reality. Also, VR interfaces do not utilize more cognitive power of the human mind, and they provide immersive user experience, thus making them feel the reality. At the same time, AR imposes various perceptual challenges as it receives the visual cues from heterogeneous multimodal sources. However, the combination of VR and AR does not increase the user workload requirement i.e., users need not lay more significant effort. On the other hand, the adult users, patients and kids may apply more workload while using extended realities and require more tolerance of discomforts [141]. Psychological consequences of humans in the virtual world must be revealed parallelly while engaging in meta life. Hence, to mitigate these consequences, the special effects of emotional intelligence have been analysed, and an improved Web-based XR framework has been suggested to improve the emotional intelligence in the metaverse life [142]. Physiological signal technology can be adapted to learn in deep about human emotions through human body signals in the metaverse ecosystem [126]. Therefore, the metaverse models should create space in between the immersive experience so that the user involvement will be disengaged from using the gadgets for a specific period of time to avoid such discomforts. Thus the amalgamation of the metaverse into Industry 5.0 presents significant challenges in human mental workload. The metaverse delivers massive volumes of data feed to specialists, such as statistics, forecasts, and suggestions. This massive amount of information on specialists could cause congestive overload and in turn will decrease the overall performance. Therefore more data accumulation will be a hindrance that must be addressed. The metaverse enables people to multitask with machines, but a technological issue or an attack could place a tremendous amount of pressure on humans.

*6) Ethical Issues:* The AR and VR devices used to access the metaverse environment in Industry 5.0 applications can capture the behaviour of the human brain and grab the intention of the user. Few applications can involve the users in gathering personal data and storing it on a permanent storage medium like blockchain. As it is a common practice for most of the users to accept the privacy policy without reading the full content, to overcome this in the metaverse environment, laws and regulations can be imposed, and ethical restrictions should be defined globally to prevent such problems. Being a next generation technology, the virtual environment should frame good moral and ethical standards to preserve a safe metaverse environment. Some of the ethical issues in the metaverse are misuse of the virtual environment, unauthorised access, incorrect information, and copyright and intellectual property violations. The regulations of the metaverse should be enhanced by formulating suitable laws that being upgraded frequently as per the industry requirements and safety concerns [4]. Ethical guidelines of conduct in the virtual world (including developers, participants/avatars, and designers) and the real world based on the principles of generic consistency for universal rights have been discussed in [143]. They have also suggested an ethical framework for the virtual world covering all virtual agents and their collaboration with real-world agents. The ethical issues concerned with the evolution and usage of gaming applications of the metaverse and Web3 were discussed in [144].

*7) Standardization:* The metaverse is a virtual environment associated with the physical world in various dimensions of Industry 5.0 applications. It is next generation Internet technology implemented using various concepts such as decentralization, submersion, automation, and proliferation. Several Industry 5.0 applications started developing these technologies with open source and standalone systems integrated with audio and visual experiences. In order to avoid facing legal issues with virtual technology, new standards and principles should be framed. The metaverse is a fantasy world where independence and liberty can lead to crimes and misconducts [145]. Proper standards must be in force to avoid negative consequences in the virtual world, or it will result in increasing the chances of abusive and criminal activities in the metaverse environment [91]. An open platform should be developed that includes a common collection of devices and methodologies where the virtual technology can be built with the available standard tools [127].

Some of the regulatory solutions for the metaverse, like standard restriction in user monitoring, emotional analysis, virtual product promotions, and simulated personas, were discussed in [146]. Furthermore, a forum for the metaverse standards (https://metaverse-standards.org/) was formed for research studies on open standards for different regulatory requirements of the metaverse ecosystem, and interoperability is being conducted. This forum is constituted by leading standardization organizations like the world wide web consortium, XR Association, and other industry leaders [147].

### B. Future Directions

Based on the challenges presented in the previous section, this section provides the feasible solutions that can address those challenges.

*1) Meta blockchain with Sixth-Generation(6G) Networks:* One of the major concerns in the metaverse ecosystem is security and data privacy, as the ecosystem itself is a virtual social world. All the threats for social media users will penetrate more in this socialized environment. Some of them are cyberbullying and cyber flashing, which is sometimes very dangerous. As the users exchange their sensitive data over autonomous networks (maybe over untrusted channels), ensuring trust among various users of the metaverse is a mandate needed for its existence. Blockchain, the distributed, immutable, tokenization, and transparent ledger technology, with its capability of intelligent resource management and efficient maintenance of stored transactions, will be an essential candidate [148].

The metaverse and the blockchain, the meta-blockchain, will ensure security and privacy-preserving transactions. However, meta-blockchain may require higher computational overhead in accompanying the virtual players and heterogeneous technologies. Therefore, 6G with higher frequencies, higher capacity, and sub-second latency response can further boost the performance of the meta-blockchain. Also, quality of experience and interactions (like holographic communication) are the significant factors in meta-blockchain, 6G will be the viable option [149], [150]. 6G networks can integrate diverse applications through their higher user data rate, terahertz frequency bands, and enabling technologies [151]. Edge computing can help in realizing faster response (0.1 ms RTT latency) and AR holographic support. Therefore, 6G with meta-blockchain can secure applications like telesurgery, digital twinning, multiplayer games, 3D printing, and many more. Furthermore, the challenges of 6G (such as access control attacks and multiplexing issues) and blockchain (scalability when transactions increase very dynamically, lack of standards and efficient decentralization systems) implementations must be considered upon its integration.

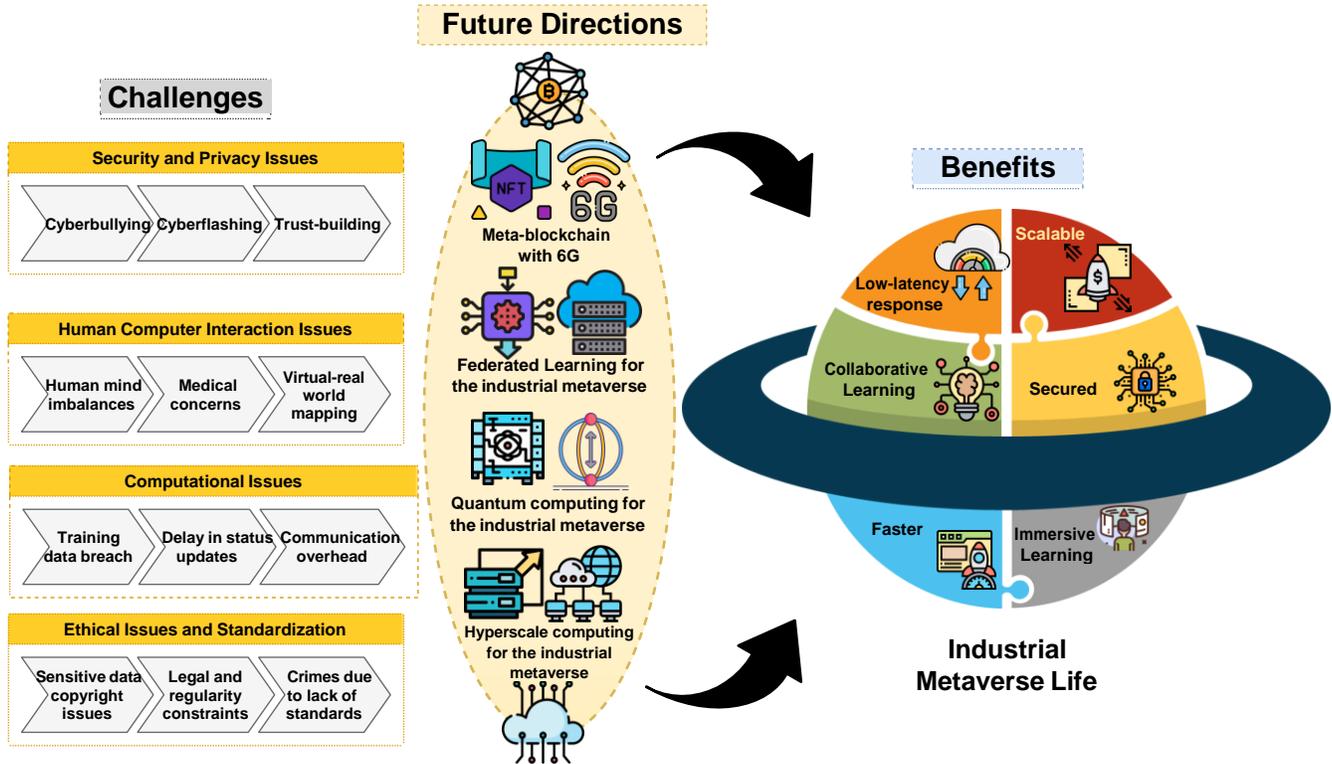

Fig. 13. Challenges, future directions and benefits of the metaverse in Industry 5.0.

*2) Federated Learning for the metaverse:* Federated learning (FL) is the collaborative learning model which trains the statistical models using multiple local data without exchanging or uploading the raw data to a centralized server. Later the encrypted parameters of the trained local model (and not data samples) are shared among themselves, and an aggregated global model is developed at the centralized server. Thus, the edge devices like smartphones can learn an aggregated global prediction model collaboratively and maintain the sensitive data at local device storage, that reduces data breaches on users' sensitive personal data. Furthermore, this shared aggregated global model is integrated with local models. Therefore, FL will ensure continual learning without data breaches due to private data aggregation and reduce the communication overhead [152], specifically for the larger models of complex scenarios with massive data [153]. Hence, FL with the metaverse can help the ecosystem to be safe and serve better. FL has been employed in various applications with massive data collection like medical institutions, AR, and digital twinning cities. The authors in [154] have proposed an FL-based solution for collaborating smart cities' digital twins ensuring faster status updates through the shared local strategy of digital twins. Here the DTs share only the parameters of the locally trained model with each other, and the global model is constructed with accumulated insights from shared local strategy.

Furthermore, blockchain can be utilized for securing the aggregated model in FL. FL has been integrated with the blockchain in many applications like IDS, resource trading, intelligent transportation, and resource allocation [155]. Therefore, FL can be employed in the meta-blockchain ecosystem for privacy-preserving data sharing. In the FL integrated meta-blockchain ecosystem, the metaverse users can train their model using local device data and share their training parameters with the model creator. This will reduce the communication overhead as the parameters are considerably smaller than the raw data. Also, the model aggregations can be carried out at the edge server to minimize the communication cost with the cloud [105].

*3) Quantum Computing in the metaverse:* Quantum computing (QC) is a phenomenon that applies principles of quantum physics to generate more powerful ways of computing. QC stores and processes the data using individual ions, atoms, photons, or electrons, creating a faster and more powerful computer. QC's two principles are quantum entanglement (lack of independence) and superposition (allow the different possible combinations of zeros and ones simultaneously). For example, QC is made of qubits, allowing the data to exist either in 1 or 0 or simultaneously in a 1 and a 0 [156]. However, the complexity lies in designing the computers for operating the world of quantum physics and QC may be a more significant threat to those applications relying on encryption [157]. Therefore blockchain can be incorporated with QC for secured computing [158]. In addition, quantum technology has been employed in communication for encrypting, and data transmission called quantum cryptography. This creates a unique quantum channel for data transmission, thus alleviating eavesdropping and other network attacks. As blockchain is one of the enabling technologies of the metaverse, quantum blockchain can make computing faster and more efficient. Also, quantum machine learning has been used in several critical applications, and the quantum-resistant security (quantum blockchain) solutions will bring the metaverse to life [159]. The metaverse ecosystem with more dynamic interactions can be implemented effectively with these security solutions.

*4) Hyperscale computing for the metaverse:* Scalability is one of the major concerns in any hyper-automation system where the number of users will be increasing every minute. Hyper-scale computing allows scaling efficiently from a minimum number of servers to thousands of servers for attaining massive scalability in the distributed computing environment. Furthermore, hyper-scale computing features horizontal scalability, redundancy, and high throughput, making it best suited for applications expecting enhanced performance, fault-tolerant, and high availability [160]. On the other hand, the metaverse, the most substantial transformation of human life, requires higher IT infrastructure facilities. Therefore, more essential edge-computing services help attain this transformation possible. The primary requirements for this transformation are hardware support processing power for the massive computing and software. Also, this massive computing transformation requires consistent (constant interaction among millions of user devices and servers), rapid response, and higher bandwidth data transfers multiple times more than the current processing capacity [161]. The hyper-scale edge data centers, with the capability to scale efficiently for accommodating the massive computing requirements of the metaverse can be a possible solution.

Therefore the metaverse ecosystem will be on life with the 6G-based meta-blockchain, federated learning, quantum blockchain, and hyper-scale computing environment.

## VI. Conclusion

This paper presents an extensive review on the metaverse applications in Industry 5.0, i.e., industrial metaverse. The metaverse can help the humans to communicate with the machines in a better way. Furthermore, it will help in bringing the human touch back into the production, help in reducing the cost of production, and realize mass personalization. In this review, the applications of the metaverse in different vertical domains of Industry 5.0, such as Society 5.0, Agriculture, Supply chain management, healthcare, smart education, disaster management, transportation, and the smart city, have been extensively discussed. Furthermore, the critical challenges in the metaverse implementation, feasible solutions, and further directions for future research are presented. From the future research perspective, several organizations such as Facebook, Neuralink, and others have been concerned with connecting the human nervous system and HCIs from input and output, respectively. Future research can focus on developing integrated chips for better HCIs to realize the potential of the metaverse in Industry 5.0. Blockchain can be utilized for secure data transactions, thereby enabling security in all aspects of the metaverse. Furthermore, federated learning, quantum computing, and hyperscale computing are expected to play a vital role in future research and development of the industrial metaverse.